\documentclass{aa}

\usepackage{txfonts}
\usepackage{textcomp}
\usepackage{tensor}
\usepackage[inline]{enumitem}

\usepackage[caption=false]{subfig}
\usepackage{booktabs}

\usepackage{tabu, multirow}

\usepackage{cmdDef}
\begin{document}

    \title{The Fornax3D project: Assembly histories of lenticular galaxies from a combined dynamical and population orbital analysis}
    \titlerunning{F3D: Assembly Histories}

    \author{%
        A.~Poci\inst{1, 3},
        R.~M.~McDermid\inst{1, 2},
        M.~Lyubenova\inst{3},
        L.~Zhu\inst{4},
        G.~van de Ven\inst{5},
        E.~Iodice\inst{6, 3},
        L.~Coccato\inst{3},
        F.~Pinna\inst{7},
        E.~M.~Corsini\inst{8, 9},
        J.~Falc{\'o}n-Barroso\inst{10, 11},
        D.~A.~Gadotti\inst{3},
        R.~J.~J.~Grand\inst{12},
        K.~Fahrion\inst{3},
        I.~Mart{\'i}n-Navarro\inst{7, 10, 11},
        M.~Sarzi\inst{13, 14}
        S.~Viaene\inst{15},
        P.~T.~de~Zeeuw\inst{16, 17}
    }
    \authorrunning{A.~Poci et al.}

    \institute{%
        Research Centre for Astronomy, Astrophysics, and Astrophotonics, Department of Physics and Astronomy, Macquarie University, NSW 2109, Australia\\\email{adriano.poci@students.mq.edu.au}\and
        ARC Centre of Excellence for All Sky Astrophysics in 3 Dimensions (ASTRO 3D), Australia\and
        European Southern Observatory, Karl-Schwarzschild-Stra{\ss}e 2, D-85748 Garching bei M{\"u}nchen, Germany\and
        Shanghai Astronomical Observatory, Chinese Academy of Sciences, 80 Nandan Road, Shanghai 200030, China\and
        Department of Astrophysics, University of Vienna, T{\"u}rkenschanzstra{\ss}e 17, 1180 Wien, Austria\and
        INAF - Astronomical Observatory of Capodimonte, Salita Moiariello 16, I-80131, Naples, Italy\and
        Max-Planck-Institut f{\"u}r Astronomie, K{\"o}nigstuhl 17, 69117 Heidelberg, Germany\and
        Dipartimento di Fisica e Astronomia `G. Galilei', Universit{\`a} di Padova, vicolo dell'Osservatorio 3, I-35122 Padova, Italy\and
        INAF - Osservatorio Astronomico di Padova, vicolo dell'Osservatorio 5, I-35122 Padova, Italy\and
        Instituto de Astrof{\'i}sica de Canarias, Calle Via L{\'a}ctea s/n, 38200 La Laguna, Tenerife, Spain\and
        Depto. Astrof{\'i}sica, Universidad de La Laguna, Calle Astrof{\'i}sico Francisco S{\'a}nchez s/n, 38206 La Laguna, Tenerife, Spain\and
        Max-Planck-Institut f{\"u}r Astrophysik, Karl-Schwarzschild-Stra{\ss}e 1, D-85748 Garching bei M{\"u}nchen, Germany\and
        Armagh Observatory and Planetarium, College Hill, Armagh BT61 9DG, UK\and
        Centre for Astrophysics Research, University of Hertfordshire, College Lane, Hatfield AL10 9AB, UK\and
        Sterrenkundig Observatorium, Universiteit Gent, Krijgslaan 281, 9000 Gent, Belgium\and
        Sterrewacht Leiden, Leiden University, Postbus 9513, 2300 RA Leiden, The Netherlands\and
        Max-Planck-Institut f{\"u}r Extraterrestrische Physik, Gie{\ss}enbachstra{\ss}e 1, 85748 Garching bei M{\"u}nchen, Germany
    }

    \date{Received XXXXX; accepted XXXXX}

    \abstract{%
    In order to assess the impact of the environment on the formation and evolution of galaxies, accurate assembly histories of such galaxies are needed. However, these measurements are observationally difficult owing to the diversity of formation paths that lead to the same present-day state of a galaxy. In this work, we apply a powerful new technique in order to observationally derive accurate assembly histories through a self-consistent combined stellar dynamical and population galaxy model. We present this approach for three edge-on lenticular galaxies from the \ftd\ project --- FCC~153, FCC~170, and FCC~177 --- in order to infer their mass assembly histories individually and in the context of the Fornax cluster. The method was tested on mock data from simulations to quantify its reliability. We find that the galaxies studied here have all been able to form dynamically-cold (intrinsic vertical velocity dispersion \(\sigma_z \lesssim 50\ \si{\kilo\metre\per\second}\)) stellar disks after cluster infall. Moreover, the pre-existing (old) high angular momentum components have retained their angular momentum (orbital circularity \(\lambda_z > 0.8\)) through to the present day. Comparing the derived assembly histories with a comparable galaxy in a low-density environment --- NGC~3115 --- we find evidence for cluster-driven suppression of stellar accretion and merging. We measured the intrinsic stellar age--velocity-dispersion relation and find that the shape of the relation is consistent with galaxies in the literature across redshift. There is tentative evidence for enhancement in the luminosity-weighted intrinsic vertical velocity dispersion due to the cluster environment. But importantly, there is an indication that metallicity may be a key driver of this relation. We finally speculate that the cluster environment is responsible for the \SZ\ morphology of these galaxies via the gradual external perturbations, or `harassment', generated within the cluster.
    }


    \maketitle


\section{Introduction}
\label{sec:intro}
Galaxy formation and evolution is the culmination of competing forces and processes over each galaxy's lifetime. These processes can be internal to the galaxy, such as the energy generated by the central super-massive black hole (SMBH) or the winds generated by star formation. They can also have external origins, such as the gravitational potential of other galaxies \citep{toomre1972} or cosmic gas filaments. Due to this `superposition' of evolutionary processes, it is difficult to isolate the impact on the galaxy from only one of them, especially when many are still occurring. The environment which a galaxy inhabits has long been suspected of altering its evolutionary path \citep{gunn1972, dressler1980, postman1984, ryden1993}, but with conflicting results on the exact impact. Field environments are relatively simple and provide a control sample for comparison with higher-density environments such as groups and clusters. This comparison is not straight forward, however, since cluster environments are a complex mixture of many, often dramatic, physical processes such as gravitational disruption (owing to the significantly deeper gravitational potential), hydrodynamic effects due to the hot intra-cluster medium (ICM), and thermodynamic effects such as shocks due to the high relative velocities that a galaxy can experience when it first encounters the ICM during in-fall.\par
A number of correlations have been observed between galactic observables and some metric for the local environment. Historically, the projected density of galaxies or \(N\)-th nearest neighbour measurements of the local density have been found to correlate with visual morphology \citep[e.g.][]{dressler1980, cappellari2011a, oh2018, gargiulo2019} and invoked to explain morphological transformations \citep[e.g.][]{bekki2002, kauffmann2004, blanton2005, donofrio2015, coccato2020}. Yet morphology has also been observed to correlate with stellar mass at fixed local density \citep{vanderwel2008}, and so the underlying cause is difficult to discern. This problem permeates through most observed correlations. Some works have shown that galaxies exhibit a lower net angular momentum for a higher local density \citep[e.g.][]{cappellari2011, cortese2019, graham2019, cole2020}, while others find that there is no additional dependence on the environment once the correlation between the angular momentum and stellar mass is accounted for \citep{brough2017}. Finally, the stellar population parameters also suffer from conflicting correlations. Some observations indicate reduced star-formation activity \citep[e.g.][]{balogh2004, poggianti2006, allen2016, owers2019}, a higher stellar metallicity \citep[e.g.][]{schaefer2019}, older stellar ages \citep[e.g.][]{thomas2005, mcdermid2015}, and a lower gas content \citep[e.g.][]{zabel2019} for higher local density, while others indicate that stellar mass is the driver instead of environment \citep{alpaslan2015, goddard2017}. Many of these correlations have also been found in recent cosmological hydrodynamical simulations \citep[e.g.][]{choi2018a, wang2018c, wang2018}. More broadly, it is not straight forward to disentangle the effects of mass and environment, and it is likely that both play a role \citep{peng2010, smith2012, mcdermid2015, wang2020b}, and joint analyses over all available parameters are needed such as those applied by \cite{christlein2005} to global galaxy properties. The morphology, mass, and other galactic properties are intricately connected through each galaxy's unique assembly history. It is therefore clear that to uncover what impact the environment has, if any, the complete assembly history must be investigated directly as a function of the environment.\par
The dynamical memory of galaxies plays an important role in attempting to disentangle such assembly histories, assisted by the (often long) dynamical times of galactic systems. As such, the stellar kinematics can provide insight into this history. Dynamical models of stellar kinematics have been employed to measure constraints on galaxy formation for a variety of morphological types and environments, based on a number of different principles. The Jeans equations have been readily applied owing to their relative simplicity and computational efficiency \citep[e.g.][and \protect\citealt{cappellari2016} for a review]{cappellari2013, watkins2013, zhu2016, zhu2016a, poci2017, bellstedt2018, nguyen2019, nitschai2020, li2020e}, though with specific assumptions about the intrinsic velocity distributions of galaxies. Distribution-function models \citep[e.g.][]{cole2017, taranu2017, pascale2018} can be quite general and computationally-efficient, but usually use parametric expressions which may not provide enough freedom. Finally, the \cite{schwarzschild1979} orbit-superposition method provides a general approach without the assumption of specific distribution functions or density distributions, while also providing a wealth of information on the intrinsic properties of the model. Though it is far more computationally-expensive, it has seen a growing diversity of applications \citep[e.g.][]{vandermarel1998, cretton1999, verolme2002, gebhardt2003, valluri2004, cappellari2006, krajnovic2009, krajnovic2015, vasiliev2013, vasiliev2019, leung2018, zhu2018, zhu2018a, vasiliev2020}. Through these models, a galaxy's merger history can be traced through the potentially-complex observed kinematics, but only when confronted with a sufficiently-sophisticated dynamical model which can access the underlying intrinsic properties \citep[e.g.][]{vandenbosch2008, lyubenova2013, krajnovic2015}. However, purely-dynamical models can not produce a chronological assembly history, since they lack information about the ages of the stars and where they might have originated.\par
This work is part of the \ftd\ survey; an observational programme to study the Fornax galaxy cluster with the Multi-Unit Spectroscopic Explorer ({\rm MUSE}) at {\rm VLT}. In total, the survey observed \(31\) members of the Fornax cluster with \(m_B < 15\ \si{mag}\), at or interior to the Virial radius \citep[\(R_{\rm vir} \sim 0.7\ \si{Mpc}\);][]{drinkwater2001}. Fornax is a well-surveyed \citep{drinkwater2001, jordan2007, davies2013, munoz2015, iodice2016, pota2018, sarzi2018, zabel2019, scott2020} galaxy cluster at a distance \(D \sim 20\ \si{\mega\parsec}\), and with a total halo mass of \(\logM[{\rm halo}] \sim 13.85\) \citep{jordan2007}. The application of the \shw\ models to \ftd\ data was showcased in \cite{sarzi2018}, and a qualitative comparison to the stellar populations was made in \cite{martin-navarro2019}. In this work, we aim to measure complete chronological assembly histories of three edge-on \SZ\ galaxies - FCC~153, FCC~170, and FCC~177 - by quantitatively combining these sophisticated dynamical modelling techniques with the measured stellar populations. They are discussed in conjunction with a previous application of this method \citep{poci2019} to a massive \([\logM[\star] \sim 11]\) field \SZ, NGC~3115, to probe any potential impact of the cluster environment.\par
This work is organised as follows: the data and target selection are briefly outlined in \cref{sec:data}, and the combined dynamical and population modelling is detailed in \cref{sec:methods}. Results for each galaxy are presented in \cref{sec:res}. The implications of these results in the context of the Fornax cluster and specific quantitative correlations are investigated in \cref{sec:discussion}.

\section{Data and targets}\label{sec:data}
\subsection{Photometry}
The photometric data for this work is taken from the Fornax Deep Survey \citep[FDS;][]{iodice2016, venhola2018}, which acquired deep photometry of the Fornax cluster out to  in the \(u\), \(g\), \(r\), and \(i\) bands using the Very Large Telescope ({\rm VLT}) Survey Telescope ({\rm VST}). We utilise the \(r\)-band photometry to model the surface brightness distribution of these galaxies. We also make use of the \(g-i\) colour to characterise the mass distribution beyond the field-of-view (FOV) of the spectroscopy (see \cref{ssec:massModel}). FDS data extend down to a surface brightness of \(\mu_r \sim 28\ \si{mag\ arcsec^{-2}}\) in the \(r\) band.\par
Distances to these galaxies were measured in \cite{blakeslee2009} via surface-brightness fluctuations. We adopt those measurements here, given in \cref{tab:masses}.
\subsection{Spectroscopy}\label{ssec:spec}
The spectral data are taken from the \ftd\ project \citep{sarzi2018}. In this work, all data products are computed on the spectral range \(\lambda \in[4600, 6700]\ \si{\angstrom}\). This range avoids the problematic sky emission lines and telluric effects. It is wide enough, however, to include many of the important absorption features for the stellar population analyses. Moreover it encapsulates the bandwidth of the \(r\) filter of {\rm VST} which is utilised in conjunction with the spectroscopy to describe the luminosity density of the stellar kinematic tracer. To prepare the data products, the data-cubes are spatially binned to a target signal-to-noise ratio \((S/N)\) of \(100\) using the Python implementation\footnote{Available at \href{https://pypi.org/project/vorbin/}{https://pypi.org/project/vorbin/}} of the Voronoi binning technique \citep{cappellari2003}. This ensures that the kinematic and stellar-population measurements can achieve measurement errors \(\lesssim 5\%\) (shown in \cref{app:schwarz}). \par
Kinematics are extracted for each binned spectrum using the {\tt pPXF} \citep{cappellari2004, cappellari2017} Python package\footnote{Available at \href{https://pypi.org/project/ppxf/}{https://pypi.org/project/ppxf/}}, which determines the line-of-sight velocity distribution (LOSVD) through moments of the Gauss-Hermite series. We extract the first six moments of the LOSVD in each bin; mean velocity \(V\), velocity dispersion \(\sigma\), skewness \(h3\), kurtosis \(h4\), and higher-order deviations \(h5\) and \(h6\). {\tt pPXF} is run with the {\rm MILES} empirical stellar library \citep{falcon-barroso2011}, and with an additive polynomial of degree \(10\) in order to accurately reproduce the line shapes. Naturally, spectra are dominated by the brightest components of the observed galaxies through the LOS, and so the extracted kinematics are effectively luminosity-weighted.\par
\begin{figure}
    \centerline{\includegraphics[width=\columnwidth]{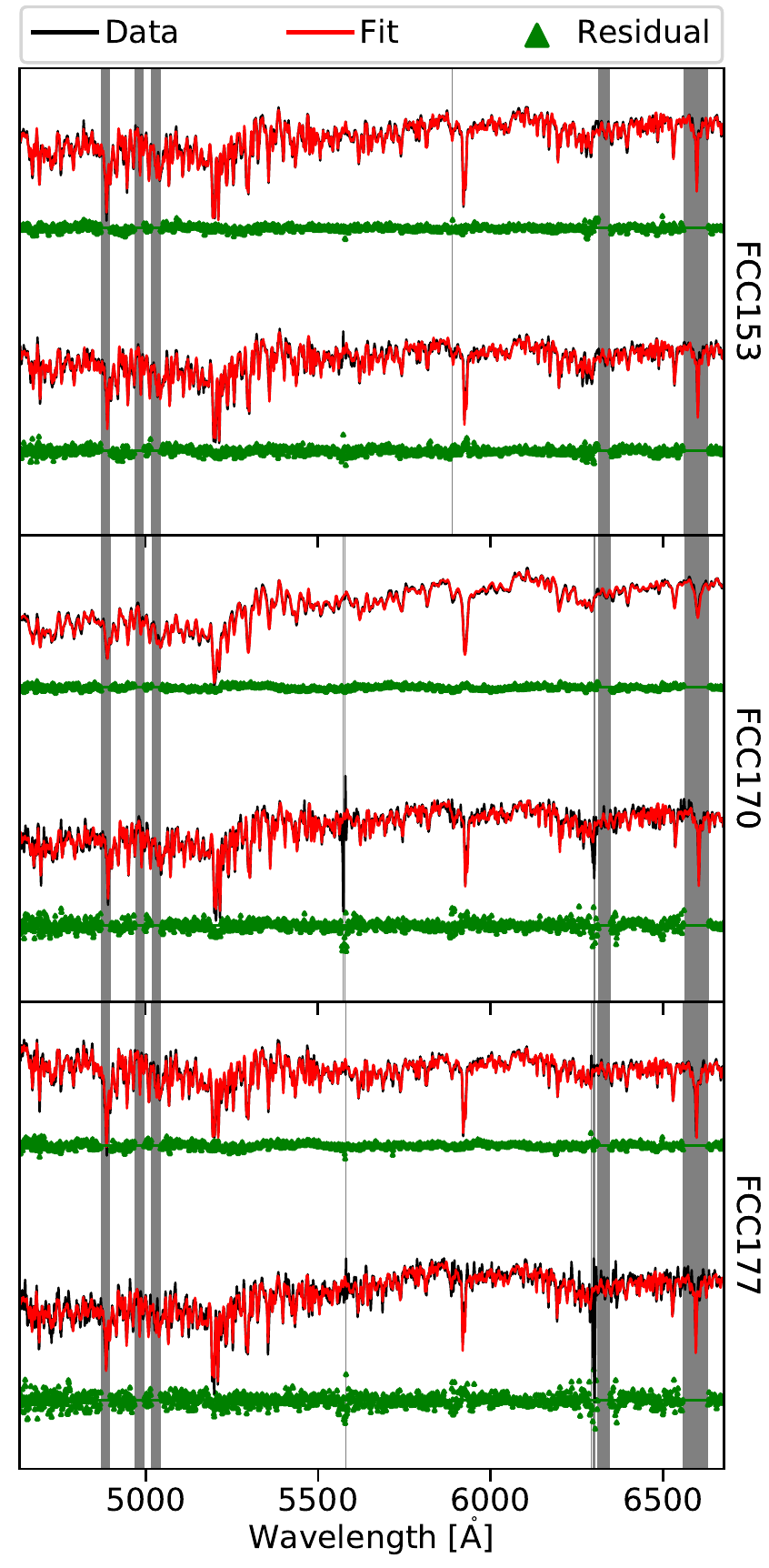}}
    \caption{Fits ({\em red}) to spectra ({\em black}) from the centre and outer regions ({\em top} and {\em bottom} of each pair of spectra, respectively) for our galaxy sample, as labelled on the right. Residuals are shown in green, offset for presentation. The grey bands show regions which are masked during the fit. All spectra are normalised, but vertically offset for presentation. It can be seen that the outer spectra are more noisy, as expected, but that in all cases the data are reproduced well by the fit.}
    \label{img:spectra}
\end{figure}

Star formation histories (SFH) and their mean stellar population properties are extracted by running {\tt pPXF} with the {\rm E-MILES} single stellar population (SSP) templates \citep{vazdekis2016} using the `BaSTI' isochrone models \citep{pietrinferni2004}. A multiplicative polynomial of degree \(10\) is included in order to account for the continuum without affecting the relative line shapes. The SSP models are normalised such that we measure luminosity-weighted stellar populations, in order to maintain consistency with the stellar kinematics and subsequent dynamical model (described in \cref{ssec:schwarz}). The stellar-population fits use a first-derivative linear regularisation with \(\Delta=1.0\), which prefers a smoother solution in the case of degeneracy between the SSP models. We assume a fixed \cite{kroupa2002} galaxy initial mass function (IMF). The canonical \cite{salpeter1955} IMF has been shown to disagree with the mass-to-light ratios from stellar dynamics \citep{lyubenova2016}, while the low central velocity dispersion of these galaxies \citep{iodice2019} is consistent with an IMF which is relatively deficient of dwarf stars \citep[e.g.][]{thomas2011, cappellari2012, wegner2012}. In this work, we explore the projected distribution of mean stellar age \((t)\) and metallicity \((\text{total metal abundance, }\chemZH)\). Representative spectral fits are presented in \cref{img:spectra}.\par
The solutions from the stellar-population run of {\tt pPXF} and the predictions from E-MILES \citep{vazdekis2010} then enable the derivation of the \(R\)-band stellar mass-to-light ratio \((\ml[{\star}]_R)\) for each spectrum, using the mass in stars and stellar remnants for the assumed IMF. This is utilised for the dynamical modelling (\cref{ssec:massModel}). We generate Monte Carlo fits to the spectra by adding random noise within the variance spectra in the data-cubes. Each spectrum is re-fit \(100\) times to generate a new distribution of SSP weights. Luminosity-weighted properties are re-derived for each weight distribution. The `uncertainty' in a given aperture is then estimated from the variance of luminosity-weighted properties across all Monte Carlo simulations in that aperture. These uncertainty maps (shown in \cref{app:schwarz}) are utilised to gauge the stability of our final results.\par
Similar data products have already been measured for these galaxies as part of \ftd\ \citep{pinna2019a, pinna2019, iodice2019}. The motivation for re-extracting them in this work is to achieve higher \(S/N\) for higher-precision stellar population parameters \citep[see, for instance,][]{asad2020} and to minimise the impact of measuring \(\sigma_{\rm los} \lesssim \sigma_{\rm inst}\) \citep[for the line-of-sight velocity dispersion and instrumental velocity resolution \(\sigma_{\rm los}\) and \(\sigma_{\rm inst}\), respectively;][]{cappellari2017}, albeit on larger spatial bins. Moreover, we fit the specific wavelength range, as discussed above. Finally, luminosity-weighted stellar populations are required for the analyses in this work as described above, while previous measurements are mass-weighted \citep{pinna2019a, pinna2019}. The new kinematics from this work are consistent with previous measurements. The luminosity-weighted ages are systematically younger than the mass-weighted determinations while the metallicities are consistent, as expected \citep{serra2007, mcdermid2015}.
\subsection{Targets}\label{ssec:target}
\begin{figure*}
    \centerline{\includegraphics[width=\textwidth]{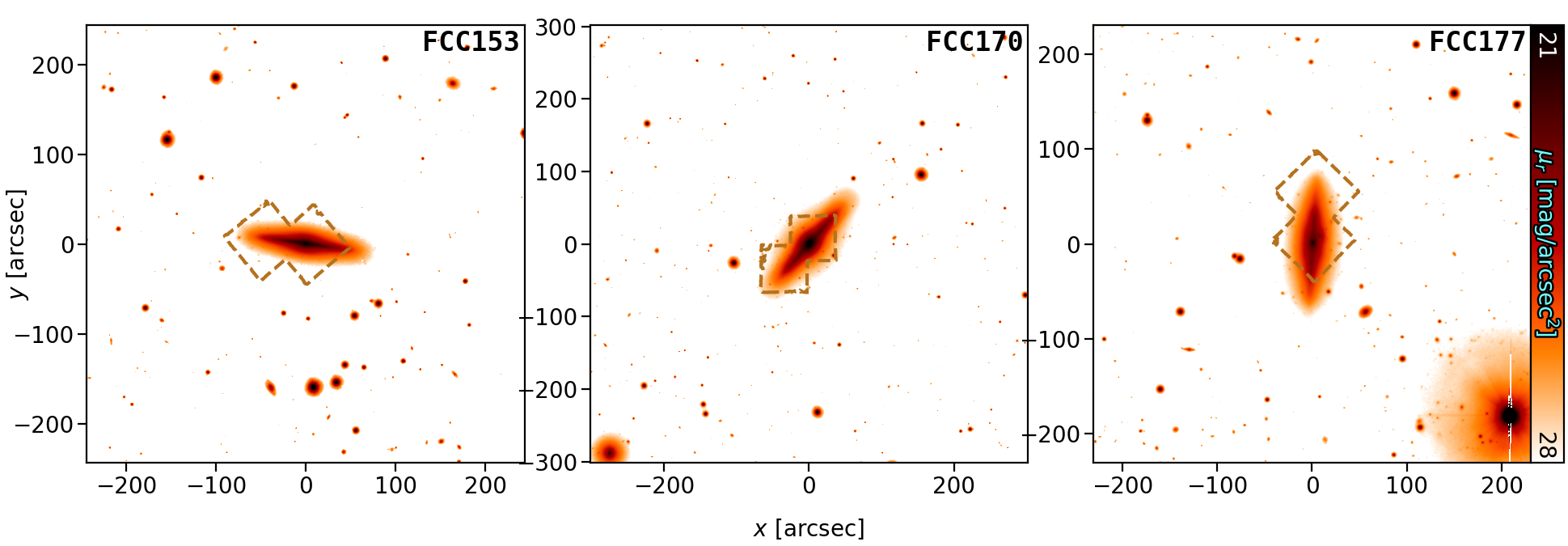}}
    \caption{Full \(r\)-band images from FDS, overlaid with the MUSE FOV in dashed brown for FCC~153 ({\em left}), FCC~170 ({\em middle}), and FCC~177 ({\em right}).}
    \label{img:photo}
\end{figure*}

For this work, due to the nature of our dynamical and population orbital analysis (\cref{sec:methods}), we selected a sub-sample of three galaxies: FCC~153, FCC~170, and FCC~177. These galaxies are all approximately edge-on, and have \SZ\ morphology. They are suitable targets for our analysis because they show no signs of dust or spiral arms. This is important because the dynamical model assumes a steady-state gravitational potential while spiral arms are transient, and dust would impact the inferences of the stellar populations. Additionally, our methodology (\cref{sec:methods}) is most robust for edge-on systems. Each galaxy has a central and outer pointing from the \ftd\ survey, ensuring that the vast majority of the stellar body is covered while retaining the high spatial resolution of {\rm MUSE}. The FDS \(r\)-band images of the three galaxies are shown in \cref{img:photo}, with the MUSE outline shown in dashed brown.
As measured from the FDS data, FCC~153, FCC~170, and FCC~177 are at a projected distance of \(1.17\si{\degree}\), \(0.42\si{\degree}\), and \(0.79\si{\degree}\) from the cluster core, respectively \citep{iodice2019b}. These galaxies have integrated \(g-i\) colours of \(0.77 \pm 0.07\), \(1.07 \pm 0.02\), and \(1.80\pm 0.03\), and \(r\)-band surface-brightness radial profiles which extend down to \(28.9\), \(29.2\), and \(29.5\ \si{mag\ arcsec^{-2}}\), respectively, as derived from the FDS photometry \citep{iodice2019b}.\par
These galaxies are the focus of the spectral analyses presented in \cite{pinna2019a, pinna2019}, where their SFH are discussed in the context of the Fornax cluster. Those works conclude that FCC~170 matured more rapidly, having plausibly evolved in an earlier group environment in the initial stages of the Fornax cluster assembly. Conversely, FCC~153 and FCC~177 are seen to exhibit relatively smooth SFH in their thin disk regions. In the study on stellar accretion fractions in members of the Fornax cluster, \cite{spavone2020} find that it is difficult to photometrically disentangle the various components of these three galaxies, since they have indistinguishable surface brightness profiles. They find that low accretion fractions \((\lesssim 50\%)\) are typical for other galaxies at cluster-centric radii similar to FCC~153 and FCC~177. Conversely, for galaxies in the region close to FCC~170, higher accretion fractions \((\gtrsim 50 \%)\) are derived. We aim, as part of this work, to place constraints on this fraction even for the galaxies which are photometrically degenerate.

\section{Stellar content of galaxies}\label{sec:methods}
We endeavour to consider the complete stellar information content available through observations. The model we fit to these data was the self-consistent combination of \shw\ orbit-superposition dynamical models, using a triaxial implementation \citep{vandeven2008, vandenbosch2008}, and stellar-population measurements derived from full spectral fitting. We employed the method described in \cite{poci2019} for this combination. We therefore refer to that work and references therein for details, but lay out the basic structure of the method and the differences with that work in this section.\par
We also ensured that our data are tracing the galaxies themselves, and not components of the cluster environment. The FDS data show that FCC~170 is within the large-scale intra-cluster light (ICL) detected towards the cluster centre \citep{iodice2019}. This ICL component was measured to have total integrated magnitudes of \(12.1 \pm 0.3\) and \(11.4 \pm 0.3\) in \(g\)- and \(r\)-band, respectively, over an area of \(\sim 432\ \si{arcmin^2}\) \citep[assuming a uniform surface brightness distribution;][]{iodice2017b}. At the distance and direction from the cluster centre to FCC~170, the ICL has a \(r\)-band surface brightness of \(\sim 27.5\ \si{mag\ arcsec^{-2}}\) \citep{iodice2017b}. In contrast, the spectroscopic data from the \ftd\ survey have a \(r\)-band target depth of \(25\ \si{mag\ arcsec^{-2}}\) in the faintest regions covered by the FOV. For our sample, the FOV extend to \(4.90\), \(5.63\), and \(9.61\ \si{\kilo\parsec}\) along the major axis for FCC~153, FCC~170, and FCC~177, respectively, at our adopted distances (see \cref{tab:masses}). In all three cases, therefore, we expect the impact of the ICL on the measured properties to be negligible, being at least \(\sim 100\) times fainter than the faintest regions of the galaxies within our spectroscopic FOV.
\subsection{Stellar mass model}\label{ssec:massModel}
One of the most crucial aspects of a dynamical model of stellar kinematics is the input mass model in which the observed tracer population resides. This is often derived from the observed photometry. We begin by fitting a multi-Gaussian Expansion \citep[MGE;][]{monnet1992, emsellem1994} to the \(r\)-band photometry from FDS using a Python implementation\footnote{Available at \href{https://pypi.org/project/mgefit/}{https://pypi.org/project/mgefit/}} \citep{cappellari2002}. This produces a projected surface-brightness model (\mgeS), which serves as the luminous tracer of the gravitational potential of the galaxy. These models are shown in \cref{app:massMGE}.\par
To reconstruct the mass, the surface brightness must be converted into surface mass density. While standard implementations of \shw\ (and indeed dynamical) models assume a spatially-constant conversion from luminosity to mass, we exploit the spatially-resolved map of stellar \(\ml[{\star}]_R\) in order to account for the prominent structures and variations in the stellar populations that are resolved by the high-quality spectroscopy. We make additional use of the deep FDS photometry to constrain the stellar populations outside of the spectroscopic FOV to constrain the dynamical model well beyond the measured kinematics. We use the predictions from the E-MILES SSP models to derive a relation between \(g-i\) colours and \(\ml[{\star}]_R\), which we assume to be of the form \(\log_{10}(\ml[{\star}]_R) \propto (g-i)\) as found empirically \citep{tortora2011, wilkins2013, mcgaugh2014, du2020d}. The smaller spectroscopic FOV, which is used where available, is thus augmented by the larger photometric FOV to generate \(\ml[{\star}]_R\) on the same extent as the photometry. While the spectroscopic measurements of \(\ml[{\star}]_R\) reach \(\lesssim 60\si{\arcsecond}\) for the three galaxies, the depth of the FDS survey allows this coverage to be extended to \(\sim 150\si{\arcsecond}\) providing a dramatic improvement to the constraints of the mass model. Using this large-scale combined (spectroscopic and photometric) \(\ml[{\star}]_R\) map, \mgeS\ is then converted to a map of surface mass density, to which a mass density MGE (\mgeT) is fit. The fits and results for all \mgeT\ are given in \cref{app:massMGE}. \cref{img:153massMGE,img:170massMGE,img:177massMGE} show deviations of up to \(\pm 30\%\) compared to a spatially-constant \(\ml[{\star}]_R\) (in projection). This approach takes into account not just these deviations in the absolute scale, but also the structures of the stellar populations, producing a more accurate mass model and subsequent dynamical model.\par
The photometric measurements of \(\ml[{\star}]_R\) are effectively `SSP-equivalent', while the spectroscopic values are derived from the full SFH. To mitigate any systematic offsets this may cause, the photometrically-derived values are re-scaled to match the spectroscopic values in the overlapping regions. We emphasise that the rapidly-varying spatial structures in the stellar populations --- caused primarily by the thin edge-on disks --- are captured by the spectroscopy, while the photometry is utilised only in the region where variations are mild. Coupled with the intrinsic symmetry of the MGE fitting, the photometric \(\ml[{\star}]_R\) serves to extend the range of the MGE model and stabilise the shape of the gravitational potential in that region. It can be seen in \cref{img:153massMGE,img:170massMGE,img:177massMGE} that there are no systematic offsets at the transition from spectroscopically- to photometrically-derived \(\ml[{\star}]_R\), and the level of noise in the colour region is no greater than the pixel-to-pixel scatter in images to which MGE is typically applied. Overall, this procedure allows for the stellar populations to be more robustly accounted for.
\subsection{\shw\ dynamical models}\label{ssec:schwarz}
The basic premise of the \shw\ method is to numerically integrate a large number of permitted orbits within a model for the gravitational potential, then measure their kinematics and compare to observations. For real observations, the gravitational potential is of course unknown and must be iteratively fit for. To achieve this, we used a triaxial implementation of the \shw\ method that has been robustly developed and validated \citep{vandenbosch2008, vandeven2008, zhu2018, zhu2018a, zhu2020, jin2019}. In this implementation, a single model is described by seven parameters: \begin{enumerate*}[label=(\emph{\alph*})] \item the three parameters describing the intrinsic shape and viewing direction of the stellar mass distribution, \(q=C/A\), \(p=B/A\), and \(u=A^\prime/A\), where \(A\), \(B\), and \(C\) are the intrinsic major, intermediate, and minor axes, respectively, and \(A^\prime\) is the projected major axis \item the mass of the central SMBH, \(M_\bullet\) \item the parameters of the dark matter (DM) profile, which is implemented as a spherical Navarro-Frenk-White (NFW) model \citep{navarro1996}. These are the concentration \(C_{\rm DM}\) and dark mass fraction at \(r_{200}\), \(f_{\rm DM}\) \item a global dynamical mass-to-light ratio, which we denote \(\Upsilon\). This parameter can shift the global depth of the potential in order to better match the observed kinematics, but does not change its shape nor therefore which orbital families can reside within it. \(\Upsilon\) is included to account for any deviations in the absolute depth of the gravitational potential due to the assumption of the IMF when computing the \(M_\star/L\) and/or systematics in the assumed DM halo model.\end{enumerate*}\par
We streamlined the search through this large parameter-space by making reasonable assumptions about some of these parameters. The masses of the SMBH were fixed according to the empirical \(M_\bullet-\sigma_\eff\) relation of \cite{kormendy2013}, using the \(\sigma_\eff\) measurements for these galaxies reported in \cite{iodice2019}. In addition, we estimated the sphere of influence \(r_i\) of each SMBH, which is utilised by the model but is not a free parameter, using the relation of \cite{vandenbosch2015}. This is expected to have little impact on the model however --- for the galaxy with the largest central velocity dispersion, FCC~170, \(r_i \approx 0.08\si{\arcsecond}\), which is below the pixel scale of MUSE. In addition, the stellar shape parameter \(u\) was fixed to \(u=(1.0-\epsilon)\) for some small number \(\epsilon\) to avoid numerical issues. This assumption is reasonable since regular fast-rotator galaxies are found to be consistent with oblate intrinsic shapes \citep{weijmans2014}. We note that mild triaxiality is still permitted in these models, with the condition that the potential must be axisymmetric in projection. The parameter-space is thus reduced to five dimensions; \(q, p, C_{\rm DM}, f_{\rm DM}, \Upsilon\).\par
Each \shw\ model corresponds to a unique intrinsic gravitational potential. The orbital families which can reside within each gravitational potential are therefore also unique. Thus, each location in the hyper-parameter-space is accompanied by its own library of numerically-integrated orbits. These orbits are characterised by the integrals of motion which they conserve, namely the binding energy \(E\), angular momentum \(I_2\), and the third non-classical conserved integral \(I_3\). Each library of orbits was generated by sampling these integrals in \((E, I_2, I_3) = (30, 20, 10)\) steps \citep[logarithmically for \(E\) and linearly for \(I_2\) and \(I_3\); see][for details of the integral sampling]{cretton2000}. The region around the best-fit model was re-computed with a higher orbit sampling of \((E, I_2, I_3) = (60, 30, 15)\) to increase the resolution of the resulting intrinsic properties. To avoid discreteness in these libraries, each orbit was dithered by a factor of \(5\), creating a cloud of orbits around each \((E, I_2, I_3)\). Using a Non-Negative Least-Squares \citep[NNLS;][]{lawson1995} fit, the model selects the best sub-set of orbits from each library which reproduces the observed kinematics in projection. It simultaneously fits a boundary constraint, which in this work is the projected luminosity distribution, such that the weights assigned to the orbits during NNLS are luminosity weights. Thus, each unique gravitational potential has a corresponding unique set of best-fit orbits.\par
By construction, \(\Upsilon\) does not change the shape of the gravitational potential. In a gravitational potential with a fixed shape but varying \(\Upsilon\), the families of orbits do not change. Rather, the velocities of these orbits are simply scaled up or down to reflect a deeper or more shallow potential, respectively, and the NNLS fit is repeated for the scaled orbits. Therefore only four parameters require the computationally-expensive numerical integration of an orbit library. The five free parameters were optimised using an adaptive grid search, whose direction and step-size depend on the existing set of evaluated models, with a large initial spread to avoid local minima. The search terminated once all surrounding models were worse fits to the data. The kinematic fits are shown in the top seven rows of \cref{img:2dmap153,img:2dmap170,img:2dmap177}. The parameter-space searches and best-fit parameters are presented in \cref{app:schwarz}.\par
To avoid artificial bias in the model due to systematic asymmetries in the data, the even (odd) kinematic moments were point-(anti)symmetrised to be consistent with the intrinsic model symmetry\footnote{using the {\tt plotbin} package, available at \href{https://pypi.org/project/plotbin/}{https://pypi.org/project/plotbin/}}. These asymmetries present deviations of up to \(\sim 6\ \si{\kilo\metre\per\second}\) in velocity and velocity dispersion with respect to the symmetrised kinematics, which is of order the measurement uncertainties on the kinematics. The `raw' un-symmetrised kinematics and their Monte Carlo-derived errors are shown in \cref{app:schwarz}.\par
\begin{figure}
    \centerline{\includegraphics[width=\columnwidth]{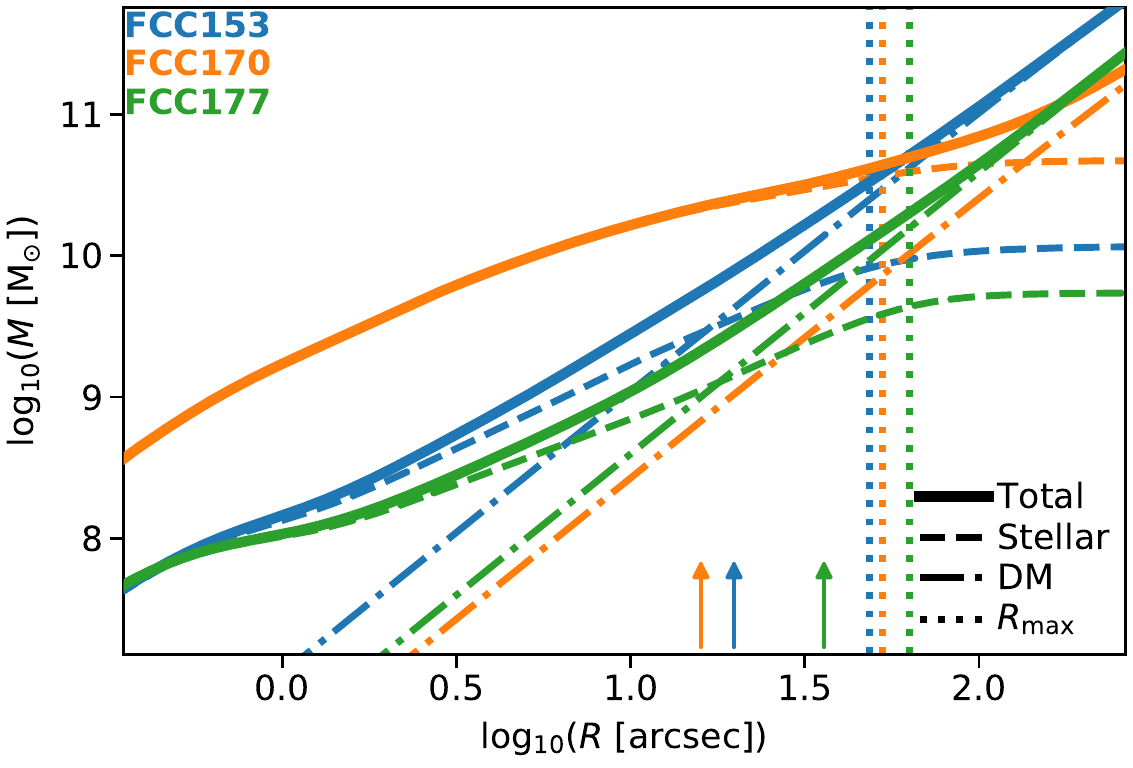}}
    \caption{Enclosed-mass profiles of the total (dynamical) mass ({\em solid line}), stellar mass ({\em dashed line}), and DM ({\em dot-dashed line}) for the three galaxies. The effective radii are denoted by the small arrows, and the radial extent of each spectroscopic FOV is shown by the vertical dotted line. The lower radial bound of the figure is set to half the width of the point-spread function from the spectroscopic observations.}
    \label{img:encm}
\end{figure}

The \shw\ models allow us to investigate the distribution of mass within these galaxies. Enclosed mass profiles are presented in \cref{img:encm}, where the maximum extent of the spectroscopy is marked by \(R_{\rm max}\), while useful quantities are provided in \cref{tab:masses}.
\begin{table*}
	\newcommand\Tstrut{\rule{0pt}{2.6ex}}         
	\newcommand\Bstrut{\rule[-0.9ex]{0pt}{0pt}}   
	\(\begin{tabu}{cc|ccc|ccc}
	    \hline\hline
	    {\rm Galaxy} & D & R_\eff & M_{\eff}^{\star} & M_{\eff}^{\rm DM} & R_{\rm enc} & M_{\rm enc}^{\star} & M_{\rm enc}^{\rm DM}\Tstrut\Bstrut\\\relax
	    & [\si{\mega\parsec}] & & [\log_{10}\si{\Msun}] & [\log_{10}\si{\Msun}] & & [\log_{10}\si{\Msun}] & [\log_{10}\si{\Msun}]\\
	    (1) & (2) & (3) & (4) & (5) & (6) & (7) & (8)\Tstrut\Bstrut\\\hline
	    \multirow{2}{*}{\rm FCC 153} & \multirow{2}{*}{20.8} & 19.80\si{\arcsecond} & \multirow{2}{*}{9.55} & \multirow{2}{*}{9.63} & 165.78\si{\arcsecond} & \multirow{2}{*}{10.06} & \multirow{2}{*}{11.43}\\
	    & & 2.00\ \si{\kilo\parsec} & & & 16.72\ \si{\kilo\parsec} & &\Bstrut\\\hline
	    \multirow{2}{*}{\rm FCC 170} & \multirow{2}{*}{21.9} & 15.90\si{\arcsecond} & \multirow{2}{*}{10.33} & \multirow{2}{*}{8.83} & 148.39\si{\arcsecond}& \multirow{2}{*}{10.67} & \multirow{2}{*}{10.74}\\
	    & & 1.69\ \si{\kilo\parsec} & & & 15.76\ \si{\kilo\parsec} & &\Bstrut\\\hline
	    \multirow{2}{*}{\rm FCC 177} & \multirow{2}{*}{20.0} & 35.90\si{\arcsecond} & \multirow{2}{*}{9.43} & \multirow{2}{*}{9.71} & 133.347\si{\arcsecond} & \multirow{2}{*}{9.73} & \multirow{2}{*}{10.84}\\
	    & & 3.48\ \si{\kilo\parsec} & & & 12.93\ \si{\kilo\parsec} & &\Bstrut\\\hline
	\end{tabu}\)
    \caption{Physical properties of the galaxy sample. \((1)\) galaxy name \((2)\) distance to the galaxy measured by \protect\cite{blakeslee2009} using surface-brightness fluctuations \((3)\) \(r\)-band effective radius taken from \protect\cite{iodice2019b} and converted into physical units at our adopted distances \((3)-(4)\) stellar and DM masses enclosed within \(R_\eff\), respectively \((5)\) radius which encloses \(98\%\) of the stellar mass \((6)-(7)\) stellar and DM masses enclosed within \(R_{\rm enc}\), respectively.}
    \label{tab:masses}
\end{table*}
It can be seen that FCC~170 is baryon-dominated within the spectroscopic FOV, while FCC~153 and FCC~177 transition to DM-dominated at or below their effective radii (given in \cref{tab:masses}). We also define \(R_{\rm enc}\), the spherical radius which encloses \(98\%\) of the stellar mass (derived by integrating the stellar mass profile). This reduces the dependence of the mass profile on the lowest surface-brightness (most uncertain) regions. These radii are in good quantitative agreement with the maximum extent of the surface brightness profiles of \cite{spavone2020}. The corresponding stellar and DM masses within \(R_{\rm enc}\) are denoted by \(M_{\rm enc}^\star\) and \(M_{\rm enc}^{\rm DM}\), respectively. These are given in \cref{tab:masses}. The amount of stellar mass outside of the spectroscopic FOV can be estimated as \(\log_{10}\left[M_\star(R=R_{\rm max})/M_\star(R=R_{\rm enc})\right]\). This gives \(0.12\), \(0.09\), and \(0.07\ \si{dex}\), for FCC~153, FCC~170, and FCC~177, respectively. While the mass in this region \((R_{\rm max} < R < R_{\rm enc})\) is not directly constrained by the kinematics, it is still constrained by the mass model described in \cref{ssec:massModel}. We explore these mass distributions further in the sections below.
\subsection{Dynamical decomposition}\label{ssec:ddec}
\begin{figure}
    \centerline{
        \includegraphics[width=\columnwidth]{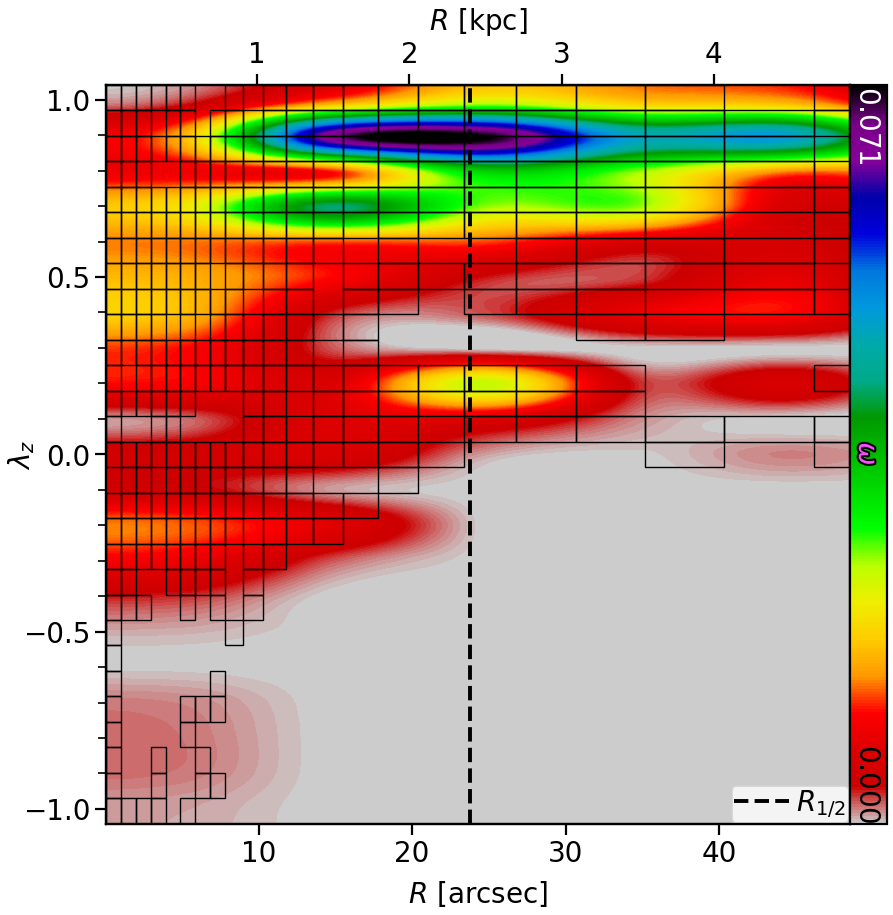}}
    \caption{Phase-space of circularity \(\lambda_z\) as a function of cylindrical radius \(R\) for the best-fit model of FCC~153. The colour represents the orbital weight from the \shw\ model, which has been normalised to an integral of unity. The dynamical decomposition is overlaid in black, where only those components which have non-zero contribution to the original model are defined. The figure is shown on the radial extent of the spectroscopy for clarity, but the decomposition is conducted over the full \shw\ model. The black dashed line is the half-mass radius, derived from \mgeT, shown for scale. The distribution indicates the prevalence of high-angular-momentum (cold disk-like) co-rotating orbits in this galaxy, with very little contribution from hot \((\lambda_z \sim 0)\) or counter-rotating \((\lambda_z < 0)\) orbits.}
    \label{img:decomp153}
\end{figure}

\begin{figure}
    \centerline{
        \includegraphics[width=\columnwidth]{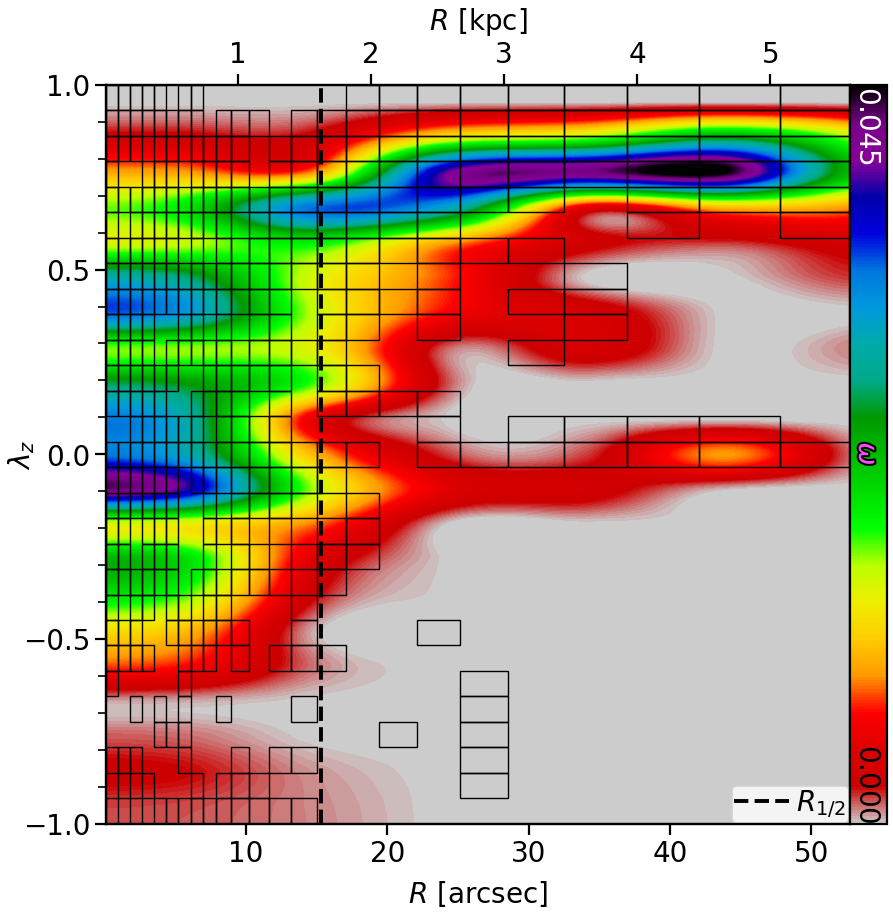}}
    \caption{Same as \protect\cref{img:decomp153}, but for FCC~170. This galaxy has a large contribution from hot central orbits, with most of the cold orbits appearing at larger radius.}
    \label{img:decomp170}
\end{figure}

\begin{figure}
    \centerline{
        \includegraphics[width=\columnwidth]{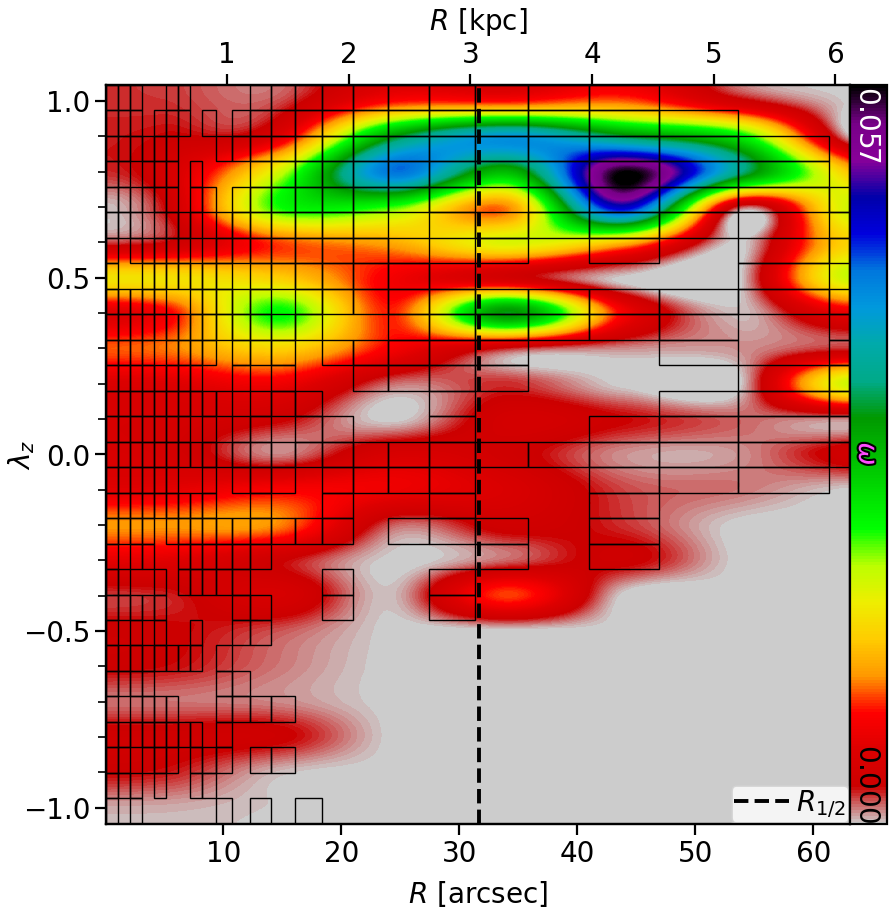}}
    \caption{Same as \protect\cref{img:decomp153}, but for FCC~177. Similarly to FCC~153, this galaxy is dominated by co-rotating cold orbits.}
    \label{img:decomp177}
\end{figure}

From the best-fit dynamical model, we used the phase-space of circularity \citep{zhu2018a}, \(\lambda_z\), and cylindrical radius, \(R\), in order to conduct a dynamical decomposition. This radius represents the time-averaged cylindrical radius of each orbit over its orbital period. The circularity is a normalised measure of the intrinsic orbital angular momentum, and we used it here to divide the \shw\ model into orbits with varying degree of rotation \((\left|\lambda_z\right| \sim 1)\) or pressure \((\lambda_z \sim 0)\) support. In order to account for the structure in the kinematics and stellar-population maps simultaneously, and motivated by tests conducted in \cite{poci2019}, we divided the phase-space into many \((\sim10^2)\) `components'. This was achieved by imposing a log-linear grid on the circularity phase-space. The radial axis was sampled logarithmically, but with a floor on the grid size. This preserves the orbital sampling from the \shw\ model\footnote{The binding energy \(E\), which is sampled logarithmically in the \shw\ models, is equivalent to the radius for a circular orbit} but avoids generating cells in the circularity phase-space which are below the spatial resolution of the data. The circularity axis was sampled linearly. This phase-space and corresponding dynamical decompositions are presented in \cref{img:decomp153,img:decomp170,img:decomp177} for FCC~153, FCC~170, and FCC~177, respectively.
This sampling in \(\lambda_z-R\) was used for all three galaxies, however the final distribution of `components' depends on the circularity distribution of each galaxy's best-fitting \shw\ model. \par
A single component is composed of a unique subset of the orbit library of its parent \shw\ model. The decomposition is an effective way of simply bundling orbits of similar properties --- in this case, angular momentum and radius. Kinematics, masses, and mass densities are computed for each component individually, based on its specific subset of orbits and those orbits' relative contribution to the original dynamical model. Thus, each component has fixed projected kinematics and spatial distributions.
\subsection{Adding stellar populations}\label{ssec:dynPop}
We now describe the extension beyond the standard \shw\ approach for the inclusion of the stellar population measurements. In order to self-consistently combine the kinematics and stellar populations, we exploited the fact that both the derivation of the SFH from full spectral fitting and the construction of the \shw\ model are based on the same principle; they are weighted integrations over many distinct populations, integrated through the line-of-sight (LOS). Specifically, the measured stellar populations are luminosity-weighted by construction as described in \cref{ssec:spec}, and the orbital weights are constrained by the surface brightness even though their dynamical properties are computed in the total gravitational potential. Therefore, we assume that the distributions of stellar and dynamical populations are the same. The weight distributions from the dynamical models were then used to derive the distributions of stellar populations that reproduce their observed maps. The result is that each orbit which contributes to the dynamical model now has an associated age and metallicity. Each dynamical component can thus be considered a mono-abundance population. By fitting age and metallicity independently, we avoided the possibility of degeneracies between them, as well as having to assume a specific age-metallicity relation. Instead, regularisation was utilised for each stellar-population fit, and is analogous to what is routinely used for spectral-fitting analyses such as in \cref{ssec:spec}. The specific implementation is detailed in \cite{poci2019}. We tested this approach using mock data from the Auriga simulations \citep{grand2016}, presented in \cref{sec:mockvalid}, and find that the main results of this work are accurate to \(\lesssim 10\%\) (\cref{img:mockCosmoDisp}). An alternative approach which uses a chemical-evolution model to derive the age-metallicity relation is presented in \cite{zhu2020}.\par
\begin{figure}
    \centerline{
        \includegraphics[width=\columnwidth]{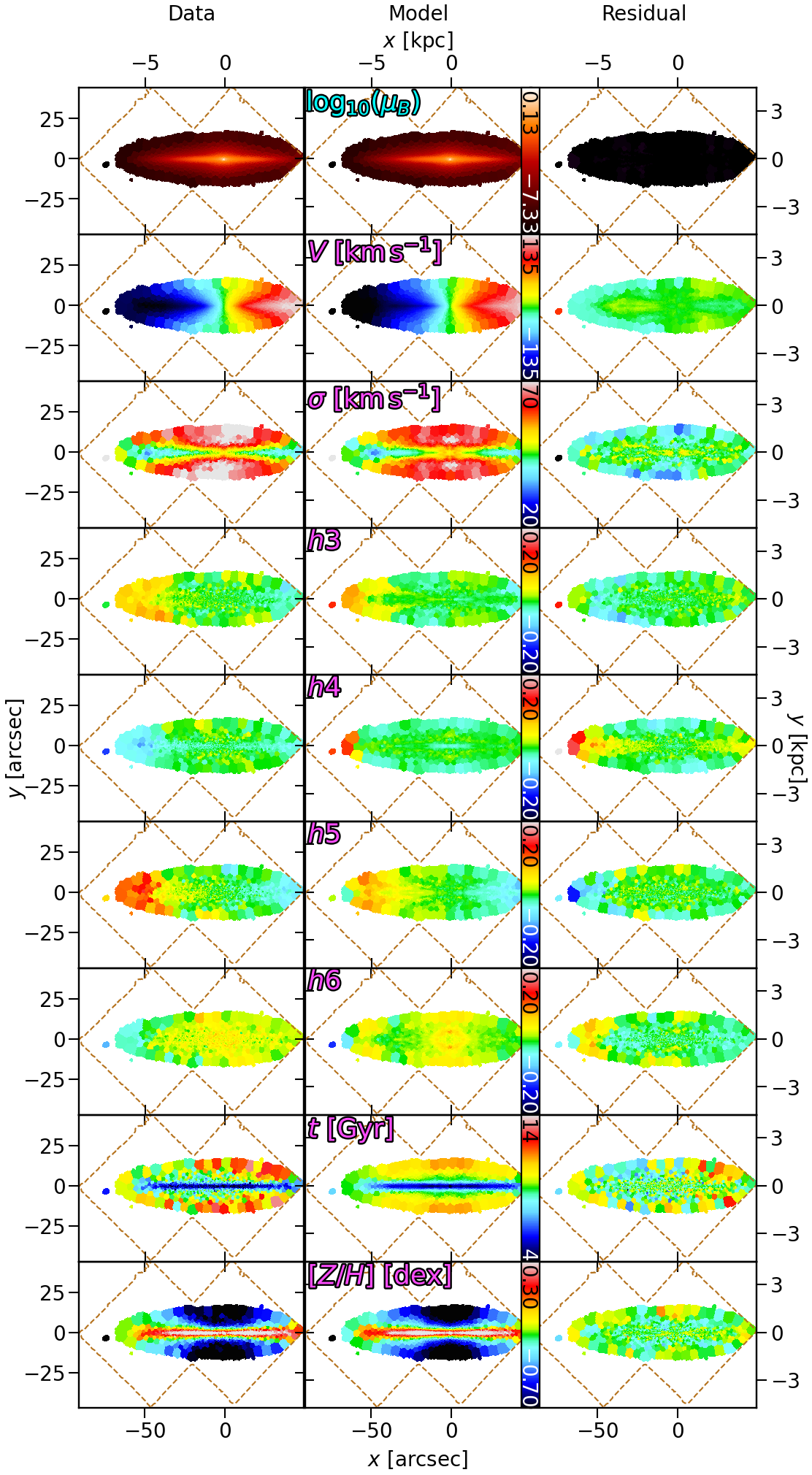}}
    \caption{Best-fitting \shw\ model for FCC~153. The data ({\em left}), fits ({\em middle}), and residuals ({\em right}) of, from top to bottom, the dynamical model (surface brightness, velocity, velocity dispersion, and \(h3-h6\)), and the subsequent stellar-population fitting (age and metallicity). The outline of the MUSE mosaic is shown in dashed brown. All residual panels show the absolute differences (data - model), but are offset such that green is zero. The stellar-population maps share a common colour-bar between galaxies for comparison. We note that for FCC~153, since the observed metallicity map reaches \(0.4\ \si{dex}\) along the major axis, and is itself an average through the LOS, we extend the upper bound for the individual components during the stellar-population fitting to \(1.0\ \si{dex}\).}
    \label{img:2dmap153}
\end{figure}
\begin{figure}
    \centerline{
        \includegraphics[width=\columnwidth]{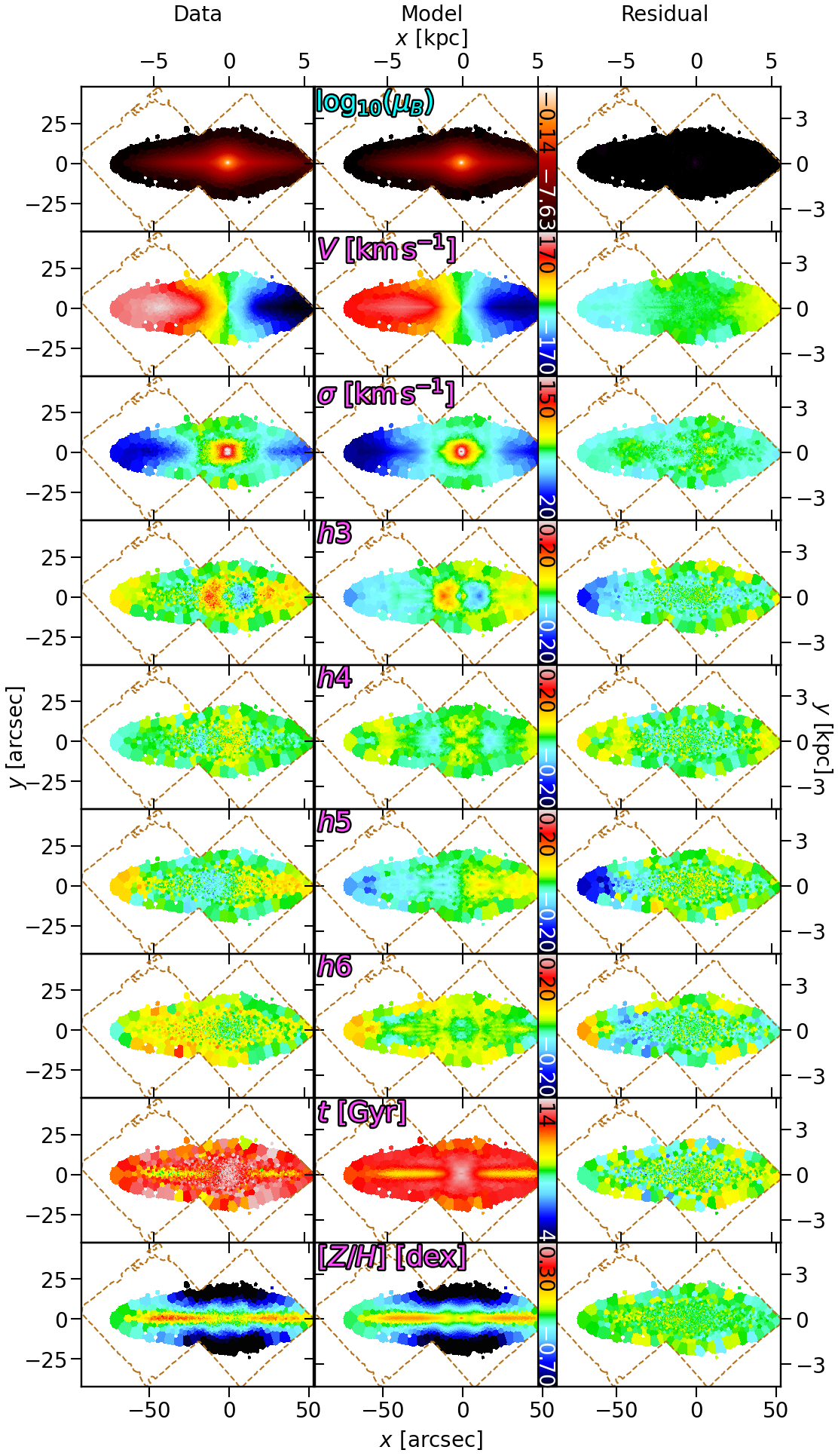}}
    \caption{Same as \protect\cref{img:2dmap153}, but for FCC~170.}
    \label{img:2dmap170}
\end{figure}
\begin{figure}
    \centerline{
        \includegraphics[width=\columnwidth]{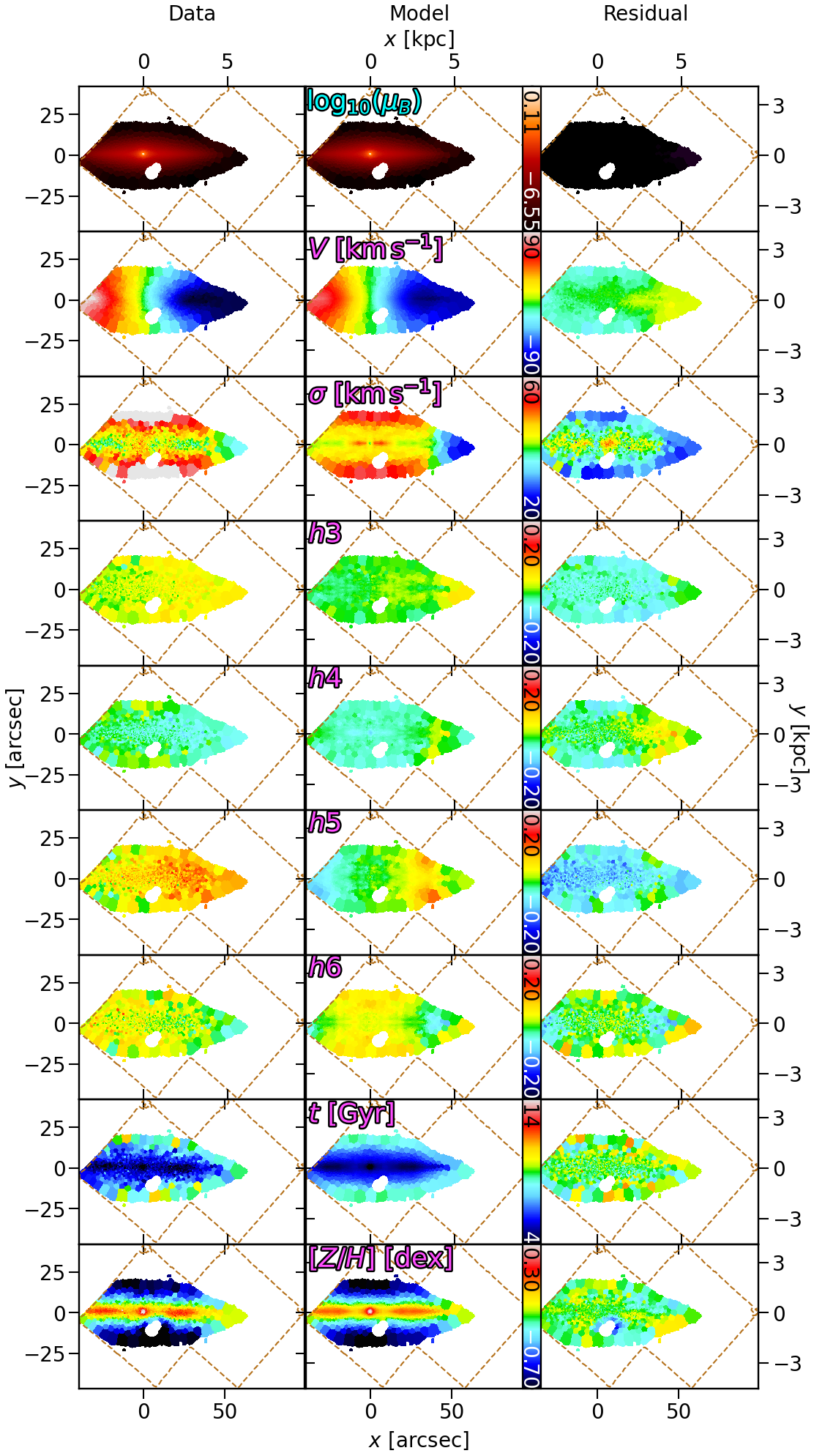}}
    \caption{Same as \protect\cref{img:2dmap153}, but for FCC~177.}
    \label{img:2dmap177}
\end{figure}

The subsequent integration through the LOS of the stellar orbits reproduces all measured kinematic and stellar-population maps. Fits to all maps are shown in \cref{img:2dmap153,img:2dmap170,img:2dmap177}.
We also conducted Monte Carlo simulations by re-fitting the stellar-population maps \(100\) times after randomly perturbing them within their measurement errors. These fits re-distribute the dynamical components in the \(t-\chemZH\) plane (without changing their kinematics), and so were used to estimate the uncertainties of our results. Using all available information -- kinematics, ages, metallicities, and density distributions -- we can now investigate the formation events that built up each galaxy.

\section{Combined dynamical and stellar populations}\label{sec:res}
\begin{figure*}
    \centerline{
        \includegraphics[width=\textwidth]{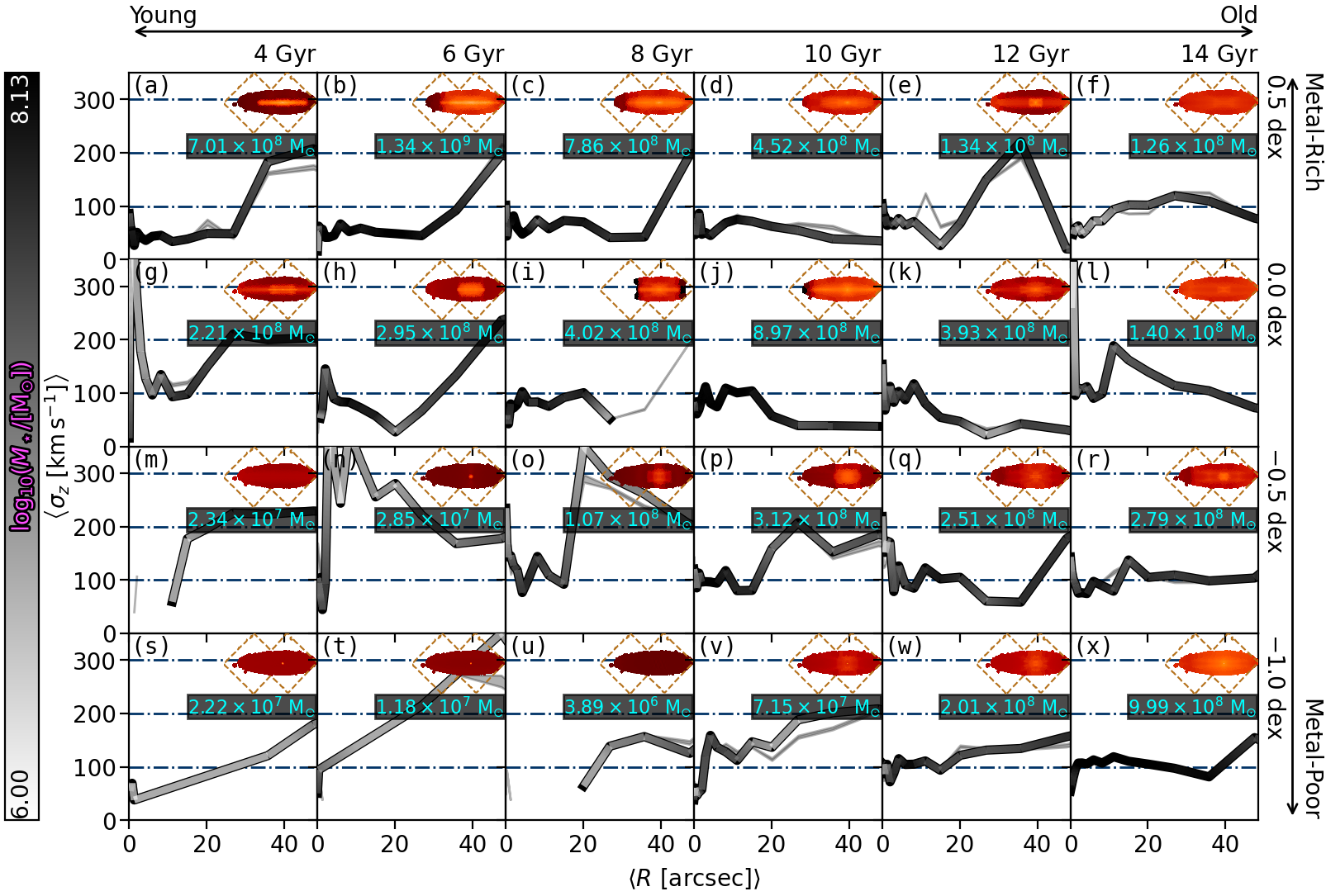}}
    \caption{Mass assembly history for FCC~153. The panels are ordered by increasing mean stellar age ({\em left to right}) and decreasing mean stellar metallicity ({\em top to bottom}). The value given at the top and right of each column and row, respectively, denotes its upper bound (inclusive). Each panel is composed of a radial profile of the vertical stellar velocity dispersion \(\sigma_z\) ({\em black/white curve}), the surface brightness distribution at the best-fitting projection ({\em top-right}) with the outline of the MUSE mosaic shown in dashed brown, and the total stellar mass within the FOV for that panel. The \(\sigma_z(R)\) profiles are coloured according to the stellar mass in that panel at that radius (sampled within the logarithmic radial bins). This indicates the spatial region in which each curve contributes most (white regions), and which regions may be impacted by numerical noise (black regions). The grey shaded regions show the spread of velocity dispersion profiles for \(100\) Monte Carlo fits to the stellar-population maps. This galaxy exhibits a dominant disk-like, metal-rich component that has steadily formed over the last \(\sim 10\ \si{\giga\year}\).}
    \label{img:mfh153}
\end{figure*}

\begin{figure*}
    \centerline{
        \includegraphics[width=\textwidth]{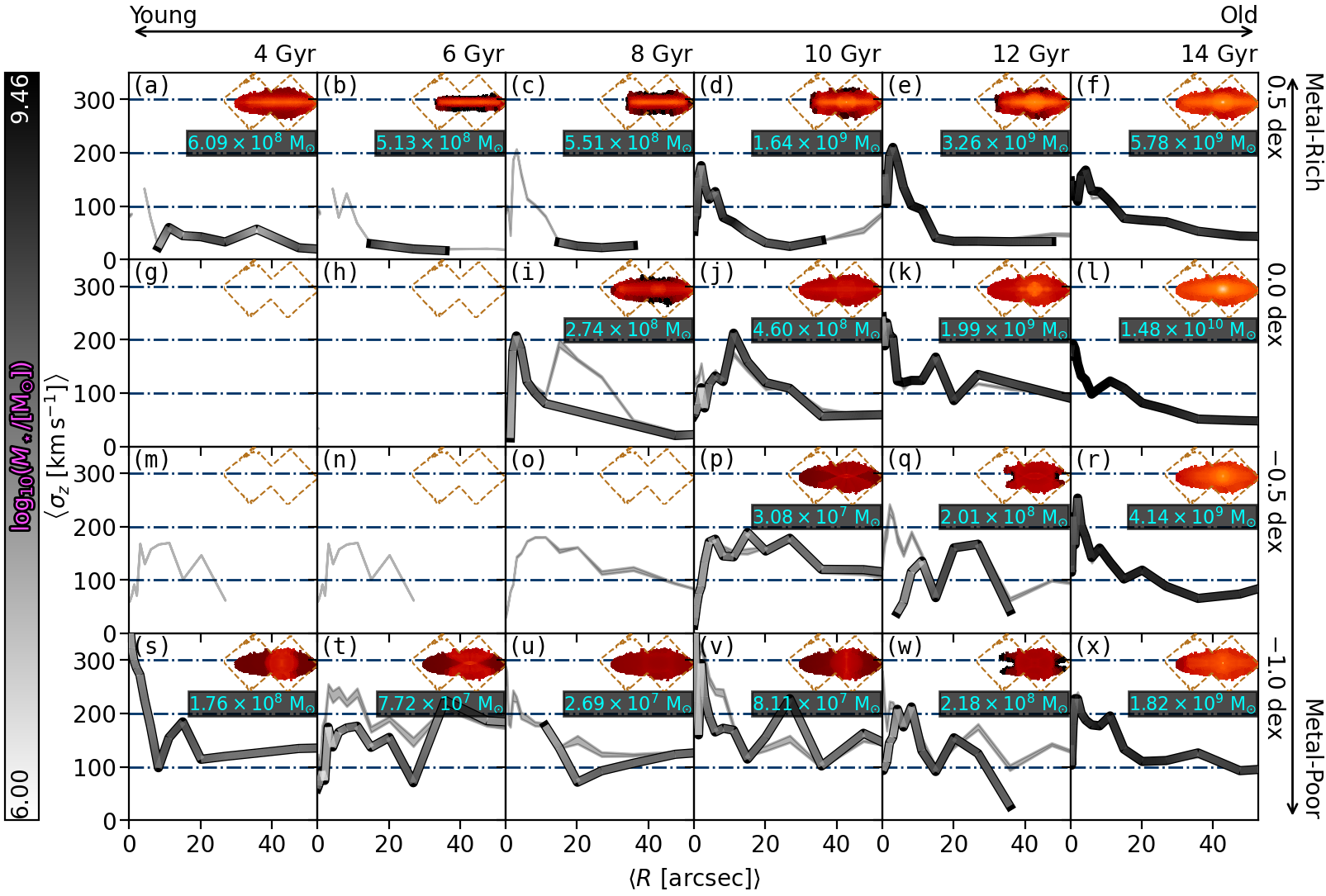}}
    \caption{Same as \protect\cref{img:mfh153}, but for FCC~170. This galaxy is dominated by an old central pressure-supported spheroidal component spanning \(\sim 1\ \si{dex}\) in metallicity. It has a secondary contribution from a progressively thinner and younger disk-like component, and a potential minor contribution from a warm metal-poor halo-like component.}
    \label{img:mfh170}
\end{figure*}

\begin{figure*}
    \centerline{
        \includegraphics[width=\textwidth]{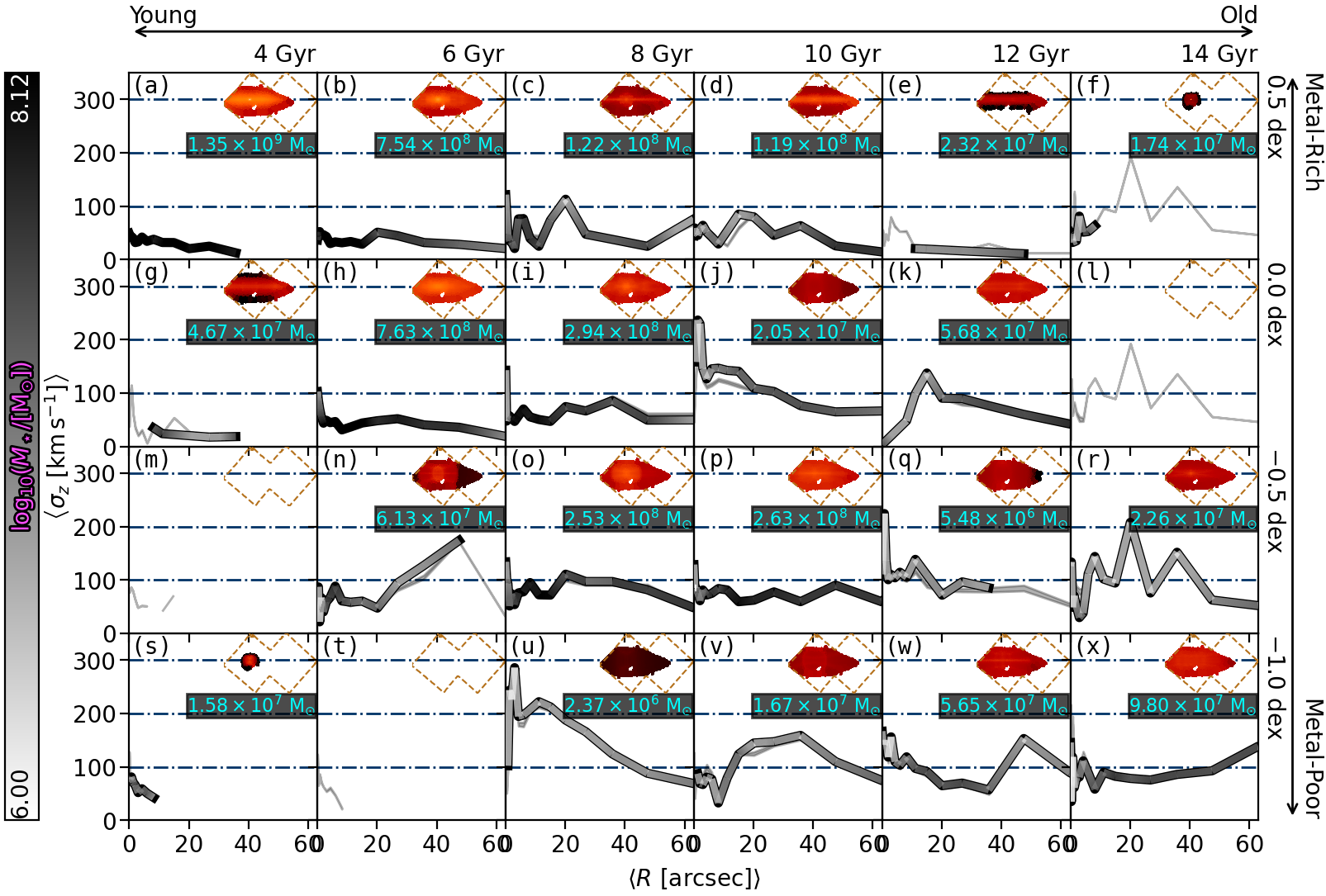}}
    \caption{Same as \protect\cref{img:mfh153}, but for FCC~177. This galaxy appears to have begun forming late. It is dominated by a young, thin disk, with contributions from dynamically-warmer and slightly older stars.}
    \label{img:mfh177}
\end{figure*}

The combination of dynamical and stellar populations is imperative to be able to decode the integrated assembly history into its constituent events. We do this using the diagnostic power of \cref{img:mfh153,img:mfh170,img:mfh177} for FCC~153, FCC~170, and FCC~177, respectively.
These figures show radial profiles of the intrinsic vertical stellar velocity dispersion \(\sigma_z(R)\) and the projected surface brightness distributions for each galaxy as a function of both age and metallicity. This combination of kinematic and population constraints effectively produces star-formation and accretion histories simultaneously, resulting in genuine mass assembly histories. The vertical velocity dispersion is a useful metric for discriminating between different dynamical structures, as well as being comparable to a variety of different observations (explored below). However, for dissecting the model into different dynamical regimes, we used the intrinsic orbital circularity to determine dynamical temperature as this property is inherently connected to the intrinsic orbital phase-space. Before exploring each galaxy individually, we first qualitatively discuss how various features of these figures are interpreted.\par
The presence of cold kinematics and flattened (`disk'-like) mass distributions are interpreted as in situ star formation, especially (though not necessarily) at high metallicity. Metal-rich and metal-poor stars in this regime would indicate that the gas likely originated from internal (recycling) and external (accretion) sources, respectively. This selection is in principle independent of age.\par
Centralised spheroidal distributions which are dynamically hot are interpreted as the in situ core or `bulge'. There is no strict selection on the stellar populations, since a large diversity has been observed in this region, especially if a stellar bar is or was present in the galaxy \citep{morelli2008, morelli2016, coelho2011, zhao2012, florido2015, seidel2015, corsini2018, barsanti2021}. Orbits at large radius with hot kinematics and with metallicities towards the metal-poor tail of the host galaxy's distribution are interpreted as the result of stellar accretion from many lower-mass systems. Such accretion is expected to be at least dynamically `warm'. This is because, although the impact of satellites may be preferentially along a particular axis \citep{shao2019}, accreted stars would nevertheless be on dynamically hotter orbits compared to the in situ cold disk. In the event of minor merging, the accreted systems will, by definition, be lower mass than the host, and via the mass-metallicity relation will thus have lower metallicities on average. Since the age of the accreted stars depends critically on the SFH of the satellites, we make no selection on age for the `accreted' stars. It is possible that some orbits in this regime have an in situ origin, from either past major mergers or significant external perturbations \citep[since low-mass accretion events themselves are not expected to perturb the existing disk significantly;][]{hopkins2008}. We nevertheless interpret this region as accretion under the assumption that it is dominated by ex situ material, subject to possible contamination by in situ material.\par
In the remainder of this section, the results are discussed briefly for each galaxy in the context of their individual assembly histories. We constrain the origin of dominant structures in each galaxy, which includes identifying the fraction of likely accreted material. Since the condition described above for the selection of accreted material favours orbits at larger radii, the limited spectroscopic FOV can bias these estimates. We instead estimate the accretion fraction as \(f_{\rm acc} = M^\star_{\rm acc}\big/M^\star_{\rm enc}\), where the accreted stellar mass \(M^\star_{\rm acc}\) is approximated for each galaxy below, and the total enclosed stellar mass \(M^\star_{\rm enc}\) is given in \cref{tab:masses}. This proposed accreted fraction is discussed further in \cref{sec:discussion}.

\subsection{FCC~153}
FCC~153 is suspected of being an `intermediate in-faller' to the Fornax cluster (\(4 < t_{\rm in-fall} < 8\ \si{\giga\year}\); as estimated from the cluster projected phase-space diagram in \citealt{iodice2019}). \cref{img:mfh153} exhibits the largest spread of metallicity of the three galaxies studied here. FCC~153 also shows the strongest recent star-formation activity, having formed the most stellar mass \((\sim 3\times 10^9\ \si{\Msun})\) in recent times \((<6\ \si{\giga\year})\), and in a kinematically-cold configuration. In fact, our model reveals that it has retained cold kinematics over all redshifts with even the oldest bins containing stars with \(\sigma_z \sim 50\ \si{\kilo\metre\per\second}\). The combination of late-time star-formation and persistent cold kinematics implies that the integrated assembly history of this galaxy (all mergers and interactions combined) has had a minimal impact, at least in the region covered by the spectroscopy. There is a suggestion of stellar accretion through the old, kinematically-warm, metal-poor population forming part of the stellar `halo'. We can use the model to estimate the mass in accreted stars by quantitatively isolating the orbits which meet the qualitative criteria discussed above. Specifically, orbits are selected with \(\chemZH \leq -0.5\ \si{dex}\) (lower half of \cref{img:mfh153}), \(\left|\lambda_z\right| \leq 0.5\), and mean guiding radius \(\overline{r} \geq 2\ \si{\kilo\parsec}\ (\sim 20\si{\arcsecond})\) to exclude any potential in situ `bulge'-like orbits (see \cref{img:decomp153}). This selection results in \(f_{\rm acc} \sim 10\%\). Under the assumption that these criteria isolate the accreted stars, we estimate an accreted mass of \(M^\star_{\rm acc} \sim 1 \times 10^{9}\ \si{\Msun}\). This selection has a luminosity-weighted average age and metallicity of \(t = 11.8\ \si{\giga\year}\) and \(\chemZH = -1.3\ \si{dex}\), respectively. Compared to the other two galaxies in this work, FCC~153 has the highest \(f_{\rm acc}\), and its relatively late in-fall may explain that. This is supported by the average age of the tentative accreted material, which implies that the main accretion events (those which dominate the luminosity-weighted average) occurred \(\lesssim 11.8\ \si{\giga\year}\) ago.

\subsection{FCC~170}
FCC~170 is believed to be an ancient in-faller to the Fornax cluster (\(t_{\rm in-fall} > \SI{8}{\giga\year}\); as estimated from the cluster projected phase-space diagram in \citealt{iodice2019}). It is the most distinct galaxy of those modelled in this work, being the most massive (\cref{tab:masses}). It is also believed to be situated closest to the cluster core. FCC~170 appears to have ceased the majority of its star formation the earliest, and its early in-fall and current position in the cluster likely played a role. The galaxy is very old, but we see evidence of recent star-formation, again in a cold configuration seemingly in spite of its environment (\cref{img:mfh170}). Overall, FCC~170 has relatively-high velocity dispersion everywhere with respect to the other two galaxies. Yet the central regions (where we probe with the spectroscopy) have remained heavily rotationally-supported over its history, with \(\sigma_z \lesssim 100\ \si{\kilo\metre\per\second}\). Applying the same accretion criteria as for FCC~153, we estimate an accretion fraction of \(f_{\rm acc} \sim 7\%\), implying \(M_{\rm acc} \sim 3\ \times 10^{9}\ \si{\Msun}\) with a luminosity-weighted average age and metallicity of \(t = 13.3\ \si{\giga\year}\) and \(\chemZH = -1.3\ \si{dex}\), respectively. With respect to FCC~153, this implies that FCC~170 experienced more accretion events of lower mass (lower metallicity).

\subsection{FCC~177}
FCC~177 is also believed to be an ancient in-faller \citep{iodice2019}, as for FCC~170. It has the lowest stellar mass and highest DM fraction of the three galaxies studied here (\cref{tab:masses}). It exhibits low velocity dispersion \((\sigma_z < 100\ \si{\kilo\metre\per\second})\) everywhere and at all times, with the younger, metal-rich populations reaching \(\sigma_z \lesssim 20\ \si{\kilo\metre\per\second}\) (\cref{img:mfh177}). We find evidence for a delayed formation, with only a small fraction of old populations \((t \gtrsim 12\ \si{\giga\year})\) and without any clear spatial structures. At later times, FCC~177 appears to have sustained modest and roughly-constant star-formation for \(t \lesssim 10\ \si{\giga\year}\). This combination of prolonged star-formation and cold kinematics is especially surprising given its early in-fall, and poses problems for the expectation of group pre-processing and cluster quenching processes. The mass budget of FCC~177 is more complicated to disentangle, especially due to the relatively diffuse mass at old ages. In fact, FCC~177 has formed the largest percentage of its stellar mass in recent times, compared to the other two galaxies. Moreover, our assembly history indicates that for lookback times greater \(10\ \si{\giga\year}\) ago, FCC~177 had just \(\logM[\star] \sim 8\), implying that the in situ component formed during that time would be lower metallicity with respect to the other two galaxies during the same period. This caveat notwithstanding, applying the same criteria as for the other galaxies, we estimate \(f_{\rm acc} \sim 6\%\). This results in \(M_{\rm acc}^\star \sim 3 \times 10^8\ \si{\Msun}\), with luminosity-weighted average age and metallicity of \(t = 12.8\ \si{\giga\year}\) and \(\chemZH = -1.5\ \si{dex}\), respectively.

\section{Mass assembly histories in context}\label{sec:discussion}
In this section we review all the evidence afforded by this technique in the context of the Fornax cluster in order to investigate the dominant processes that built up the stellar mass in these galaxies. By analysing the trends in \cref{img:mfh153,img:mfh170,img:mfh177}, and exploring them more quantitatively throughout this section, we can constrain certain formation mechanisms.\par
Interestingly, we see a diversity in the assembly histories of the three galaxies studied here via the different distributions of mass between \cref{img:mfh153,img:mfh170,img:mfh177}. Yet the persistence of kinematically-cold orbits is common throughout all of the galaxies for all stellar ages. The observation of such kinematics for old populations places constraints on both internal and external disruption processes. Owing to the archaeological nature of the methodology employed here, all stars are observed in their present-day, not formation, configurations. It is clear, therefore, that in order for these orbits to remain kinematically-cold, those stars need to not only form as such, but also experience little-to-no subsequent disruption until the epoch of observation. This implies that neither internal instabilities nor the cluster potential (or other members) can cause significant perturbations to the kinematics of the central regions of these galaxies (though this is discussed further in \cref{ssec:angmom,ssec:sz}). For the same reason, we argue that these galaxies have likely not experienced any high-mass-ratio mergers, as they would have similarly disrupted these old cold orbits \citep{hopkins2008}.\par
There seems to be no lack of historic star-formation activity in these galaxies. This is perhaps most surprising for FCC~170 which exhibits by far the oldest mean stellar age, and is purported to reside in the central region of the cluster. We find evidence for the continued formation of stars in all three galaxies down to relatively young ages, and at super-solar metallicity. These episodes occurred comfortably after each galaxy is suspected to have entered the cluster. Their metallicity is consistent with self-enrichment, and thus in conjunction with their kinematics, these stars very likely formed in-situ from recycled gas.\par
The accretion of low-mass stellar systems is expected to deposit material into the outer stellar `halo' regions of galaxies. It is also expected to contribute significantly to the present-day stellar mass of galaxies \citep{oser2010}. We have estimated, however, low accretion fractions \((< 10\%)\) for the galaxies studied here. There are two sources of uncertainty in the \(f_{\rm acc}\) estimates in this work; contamination by in situ stars in the region we consider `accreted', and excluding some accreted material which resides at lower radius. We can not strictly exclude an in situ contribution to these accretion fractions, but such contamination would imply an intrinsic accreted fraction even lower than estimated here. Without major mergers, any in situ stars that satisfy the proposed criteria for accretion are difficult to explain, unless external perturbations from the cluster have caused dramatic transformations. Moreover, \cite{karademir2019} find that mergers with smaller mass ratios deposit stars at larger radii. Once again, since we have argued against major (or even a significant amount of minor) mergers, it is plausible that at least the majority of accretion for these galaxies resides at large radius. \cite{davison2020} similarly find that for galaxies in the {\rm EAGLE} simulation, most of the accreted mass is deposited beyond the half-mass radius \(r_{1/2}\) for host stellar masses within the range of our Fornax galaxies. We nevertheless caution that \(f_{\rm acc}\) is subject to these uncertainties, and highlight that the other main conclusions of this work do not depend on the measurements of accretion. While the mass models are constrained over the full extent of the galaxies using the FDS photometry (\cref{ssec:massModel}), we can not exclude higher accretion fractions being found at larger radii as inferred by, for instance, \cite{pulsoni2018} for the stellar mass range probed by our sample.\par
A lack of accretion can be explained by the high relative motions of member galaxies within a cluster, and reduced merging has been seen previously for cluster members with respect to the field \citep{berrier2008, pipino2014}. Specifically for the Fornax cluster, its members and their globular cluster (GC) populations have been analysed previously \citep{jordan2015, fahrion2020}. \cite{fahrion2020} finds that FCC~170 has a significantly reduced number of GC for its stellar mass, and those that is has are notably metal-poor. While the numbers of GC for FCC~153 and FCC~177 are less unusual, since the hosts are themselves lower stellar mass, their GC are also more metal-poor compared to the stellar body by \(\sim 1\ \si{dex}\). This implies that the GC originated in lower-mass systems. Once again, this suggests a lack of major mergers, and low incidence of minor merging for the three galaxies studied here. Low accretion fractions for these three galaxies were also inferred from the analysis of \cite{pinna2019a}. These galaxies appear to have been shut off from sources of external material by the cluster environment, which has likely stifled their growth. Their stellar mass assembly was able to continue through in situ star formation, but has ceased in the present day likely due to the exhaustion of internal gas in conjunction with the lack of replenishment.

\subsection{The stellar age--velocity-dispersion relation}\label{ssec:angmom}
Here we quantitatively explore some of the correlations alluded to in the assembly histories. To this end, we investigate the vertical component of the intrinsic stellar velocity dispersion, \(\sigma_z\), as a function of formation time of the stars \citep[converted to redshift assuming the cosmology of][as implemented in {\tt astropy}]{ade2016}. This stellar age--velocity-dispersion relation (AVR) has been studied previously in the Local Group \citep{wielen1977, nordstrom2004, rocha-pinto2004, seabroke2007, martig2014, sharma2014, beasley2015, hayden2017, grieves2018, bhattacharya2019, mackereth2019} and a number of cosmological and idealised simulations \citep{bird2013, aumer2016, grand2016, kumamoto2017}. The gas-phase AVR is also well-studied \cite[e.g.][]{wisnioski2015}. While the stellar AVR is derived through the properties of stars of different ages within individual galaxies, in contrast, the gas-phase AVR is measured via the global properties of different galaxies observed directly at different redshifts. In all cases, \(\sigma_z\) is seen to decrease towards the present day, but with competing explanations as to the physical driver of this relation. It is often thought to be the result of either internal instabilities whose cumulative effects have disturbed older stars more \citep{saha2010, aumer2016, grand2016, yu2018} or that populations of stars which formed at high redshift inherited higher random motion from their surroundings, which has been decreasing towards the present day as conditions stabilise \citep{noguchi1998, bournaud2009, bird2013, leaman2017, ma2017}. Since the AVR pertains to the conditions which lead to star-formation, the measurements of these properties are restricted to the disk plane, as this is where in situ star-formation is expected to occur.\par
\begin{figure}
    \centerline{
        \includegraphics[width=\columnwidth]{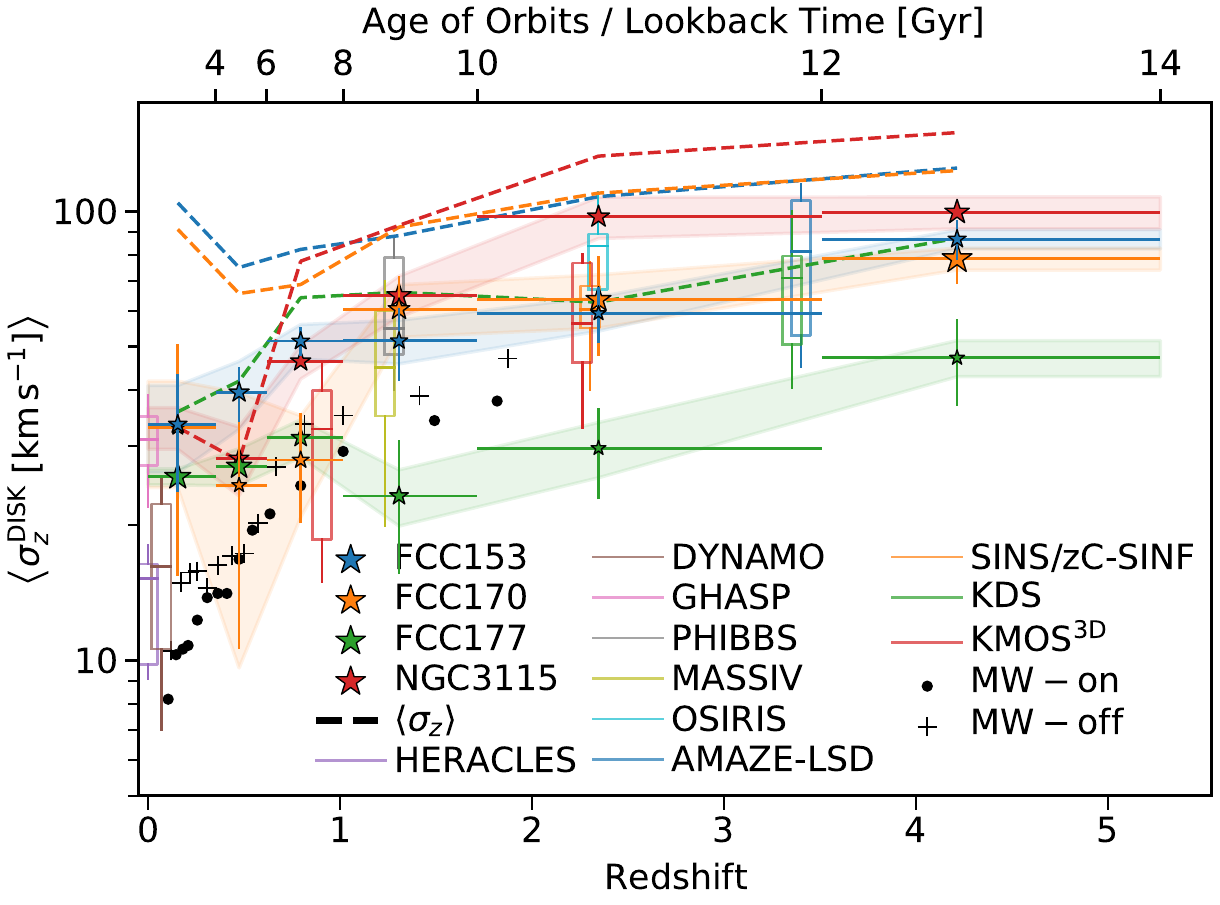}
    }
    \caption{Stellar disk AVR as derived from our models. The coloured stars are the galaxies modelled in this work \protect\citep[and][for NGC~3115]{poci2019}. The symbol size is proportional to the fractional stellar mass in each age bin, for each galaxy independently. The horizontal error-bars denote the width of the age bin. The vertical error bars are computed as the weighted standard deviation within each age bin, for the best-fit model. The shaded regions show the spread in \(\sigma_z\) for \(100\) Monte Carlo fits to the stellar-population maps. The dashed curves show the stellar AVR of the four \SZ\ galaxies when all orbits are included (no selection on orbital circularity). The box-whisker plots are literature measurements of cold gas disks, from {\rm HERACLES} \protect\citep{leroy2009}, {\rm DYNAMO} \protect\citep{green2014}, {\rm GHASP} \protect\citep{epinat2010}, {\rm PHIBBS} \protect\citep{tacconi2013}, {\rm MASSIV} \protect\citep{epinat2012}, {\rm OSIRIS} \protect\citep{law2009}, {\rm AMAZE-LSD} \protect\citep{gnerucci2011}, {\rm SINS} \protect\citep{schreiber2009} and {\rm zC-SINF} \protect\citep{schreiber2014}, {\rm KMOS}\(^{3{\rm D}}\) \protect\citep{wisnioski2015}, and {\rm KDS} \protect\citep{turner2017}. The black dots and crosses are Milky-Way stellar measurements \protect\citep{yu2018} for stars on \((|z| < 270\ \si{\parsec})\) and off \((|z| > 270\ \si{\parsec})\) the plane, respectively. Galaxy disks become dynamically colder towards the present day. The cluster \SZ\ galaxies have a higher contribution from warmer orbits at more recent times compared to the field galaxy (comparing the full and disk-only \(\sigma_z\)). The Milky-Way, despite its higher stellar mass, is dynamically colder than the \SZ\ galaxies studied here.}
    \label{img:cosmoDisp}
\end{figure}

The stellar AVR measured in this work for the three Fornax galaxies are presented in \cref{img:cosmoDisp}, with comparisons to literature measurements.
We track the velocity dispersion as a function of formation redshift, marginalised over metallicity and radius. That is, at fixed age, the metallicities are averaged according to their luminosity-weighted contribution to the model, then similarly for radius. This maintains the appropriate weighting such that the final \(\sigma_z\) measurements are also luminosity-weighted. For consistency with literature measurements, we measure these properties for the `disk'-like orbits of the models; that is, we consider only those orbits with \(\left|\lambda_z\right| \geq 0.8\). The resulting relations are shown by the large stars in \cref{img:cosmoDisp}. Additionally, the relations for all orbits (with no selection on circularity) are given by the dashed curves.\par
All of these galaxies show the same general trend of decreasing \(\sigma_z\) with decreasing stellar age. The trends we measure for the stellar \(\sigma_z\) are flatter than those for direct gas measurements, as the stars do not reach the coldest dynamical temperatures observed in gas in the present day. This is in agreement with predictions from simulations \citep{pillepich2019}. We also see that the field galaxy NGC~3115, despite being two times more massive than our most massive Fornax object, exhibits comparable vertical velocity dispersion. By comparing the disk AVR to the full-orbit AVR, it can be seen that all three Fornax galaxies show a measurable contribution from non-disk-like orbits at all ages. Together, these observations imply that we are likely measuring the impact of the cluster on the dynamics of its galaxies due to so-called `harassment' \citep{moore1996}; frequent though minor gravitational interactions between galaxies in close proximity. External perturbations have also been seen to cause heating coincident with the time of the interaction \citep{grand2016}.\par
Our results are also compared in \cref{img:cosmoDisp} to data from the Milky-Way, both on \((|z| < 270\ \si{\parsec})\) and off \((|z| > 270\ \si{\parsec})\) the disk plane \citep{yu2018}. For comparison, the physical pixel scale of the data used in this work is \(\sim 20\ \si{\parsec\per pixel}\), but with a real physical resolution of \(\sim 70\ \si{\parsec}\) due to the point-spread function of the observations. So while our models are not separated based on height above the plane, they still probe the most dynamically-cold physical scales, meaning that any differences between these results is not due to spatial resolution effects. All four \SZ\ galaxies exhibit a systematic increase of \(\sigma_z\) with respect to the Milky-Way, which is expected since galaxies that can support spiral arms should be dynamically colder. In this case, the offset is likely a combination of the different morphology and environment, yet the general shape of the relation is preserved despite these differences.\par
As discussed above, each galaxy retains a significant portion of mass with cold kinematics and disk-like morphology at the oldest age. Specifically, these oldest age bins (as seen in the present day) exhibit \(\sigma_z \sim 50\ \si{\kilo\metre\per\second}\) on the disk plane at intermediate radii and high metallicity. This is inconsistent with internal heating whose effect should be maximal for the oldest stars. For instance, the simulations of \cite{aumer2016} show that for the oldest stars, internal heating will increase \(\sigma_z\) by \(\sim 15-20\ \si{\kilo\metre\per\second}\) above the value at birth. For the old stars we measure in the present day which have \(\sigma_z \sim 50\ \si{\kilo\metre\per\second}\), this would imply that they were born with \(\sigma_z \sim 30-35\ \si{\kilo\metre\per\second}\) at \(z=4-5\), which is significantly lower than the gas measurements at that epoch. Therefore, we conclude that the AVR for these galaxies is the result of hotter dynamical temperatures at early times, while further minor heating is contributed by the cluster interactions.\par

\begin{figure}
    \centerline{
        \includegraphics[width=\columnwidth]{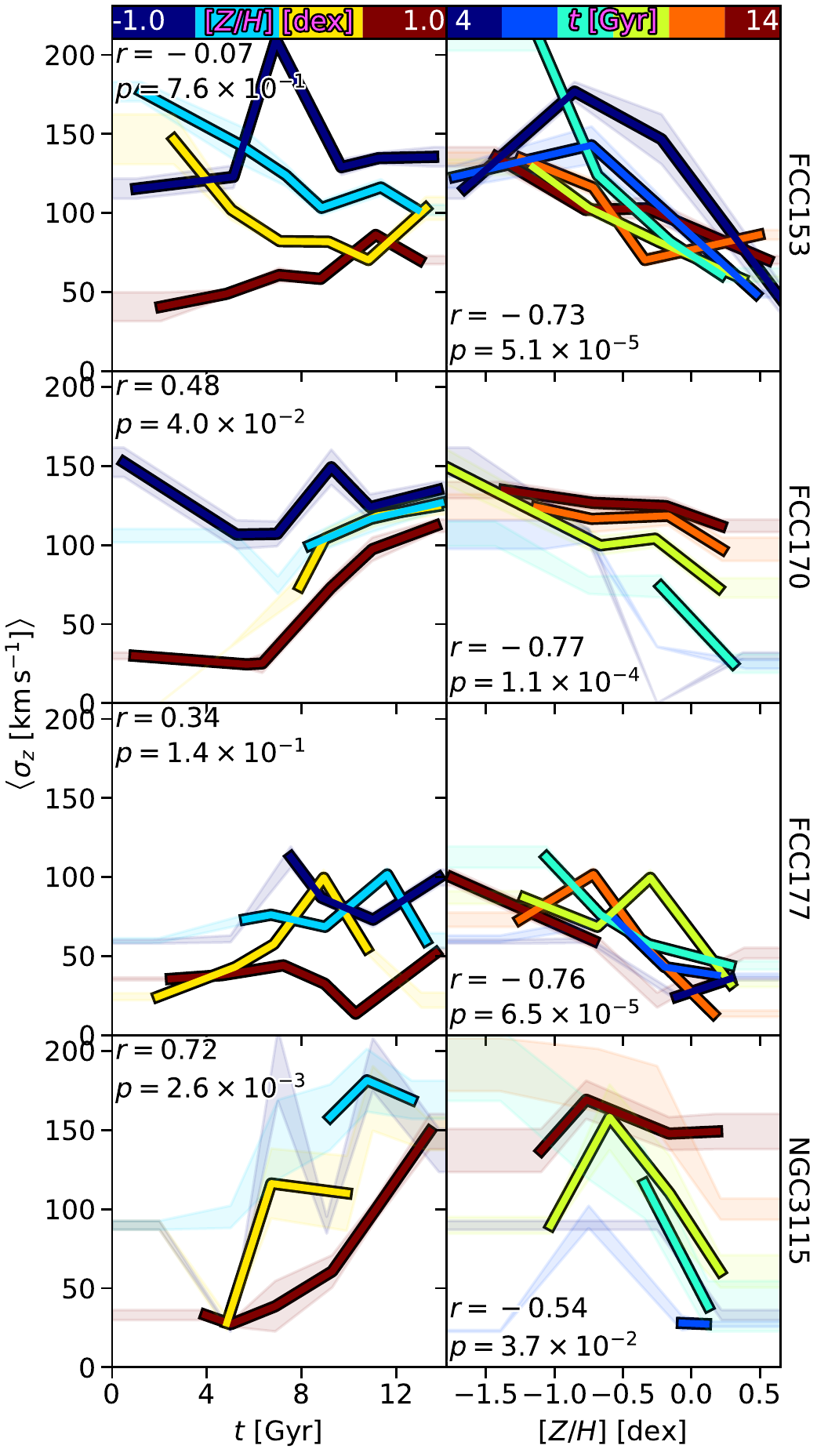}
    }
    \caption{Correlations of \(\sigma_z\) with stellar age at fixed metallicity ({\em left}), and stellar metallicity at fixed age ({\em right}) for, from top to bottom, the three galaxies studied in this work, and the field \SZ\ NGC~3115 \protect\citep{poci2019}. The curves are coloured by their age/metallicity bin corresponding to those of \cref{img:mfh153,img:mfh170,img:mfh177}. The Spearman rank coefficient \(r\) and the associated \(p\)-value, computed for all curves collectively in a given panel, are inset. The shaded regions correspond to the variations derived from \(100\) Monte Carlo fits to the stellar population maps. The stellar AVR at fixed metallicity exhibits lower significance than the \(\chemZH-\sigma_z\) at fixed age.}
    \label{img:fixedSPDisp}
\end{figure}

Closer inspection of \cref{img:mfh153,img:mfh170,img:mfh177} indicates a \ND{3} correlation between mean \(\sigma_z\), stellar age, and stellar metallicity, yet the results of \cref{img:cosmoDisp} marginalise over metallicity. We therefore compute these relations without such marginalisation, to investigate the impact of age and metallicity independently. These are presented in \cref{img:fixedSPDisp}.
In order to avoid numerical noise which could be introduced through the increasingly-complex selection criteria, we conduct this analysis on the full diversity of orbits (without selecting on circularity). The curves in \cref{img:fixedSPDisp} are constructed by collecting individual rows and columns of \cref{img:mfh153,img:mfh170,img:mfh177}. Each panel of those figures is integrated along the radial profile, preserving the luminosity weighting at each point, to produce a single \(\sigma_z\) measurement for that panel. Each row in the assembly histories corresponds to a single curve of the AVR at fixed metallicity (left column of \cref{img:fixedSPDisp}), while each column in the assembly histories corresponds to a single curve of the \(\chemZH-\sigma_z\) relation at fixed age (right column of \cref{img:fixedSPDisp}). The Spearman rank correlation coefficient \(r\), which indicates the strength and direction of a trend, is computed using the {\tt scipy} implementation for all curves in a given panel. The corresponding \(p\)-value is also shown for each panel, which indicates the probability that the two axes are uncorrelated.\par
We observe a significant \(\chemZH-\sigma_z\) correlation at fixed age, such that the more metal-poor stars are dynamically hotter. Similar correlations between the metallicity and vertical velocity dispersion have been seen previously for the Milky-Way \citep[for the iron abundance \(\chemFeH\), and typically with non-trivial selection functions;][]{meusinger1991, ness2013a, minchev2014, grieves2018, arentsen2020} and for {\rm M 31} \citep{dorman2015} but those results are marginalised over age. Similarly, all previous studies of the stellar AVR have been marginalised over metallicity --- with the exception of \cite{sharma2020}, discussed below. Interestingly, \cite{guiglion2015} see the inverse trend of \(\sigma_z\) with \(\chemMgFe\) at fixed \(\chemFeH\) for the Milky-Way. \cref{img:fixedSPDisp} shows that the AVR is a weak correlation once metallicity is accounted for, quantified by the correlation coefficients in each case. At fixed age, the \(\chemZH-\sigma_z\) relation is significantly more correlated than the AVR at fixed metallicity. We emphasise that the stellar AVR in \cref{img:cosmoDisp} (even the dashed full-orbit curves) exhibits a correlation which is consistent with previous measurements when metallicity is not taken into account. This means that the result in \cref{img:fixedSPDisp} can not be due to any degeneracy between age and metallicity in our models. Furthermore, age and metallicity are fit independently in \cref{ssec:dynPop}, and the spatial coherence of the dynamical components (each spatial bin is not independent) is exploited to reduce possible degeneracies within each fit. At face value, this result implies that \(\chemZH-\sigma_z\) is the underlying physical correlation, while the impact of age (or formation redshift) is of secondary importance. In this scenario, the stellar AVR would manifest through the age-metallicity relation and its scatter. Finally, while the results in \cref{img:fixedSPDisp} include the full diversity of orbits from our models, we confirmed, by removing the suspected accretion components (via the same selection identified in \cref{sec:res}) that neither the direction nor the relative significance of these correlations change. This implies that the results of \cref{img:fixedSPDisp} are not merely driven by the fact that accreted material is often dynamically hotter and relatively metal-poor, but rather that it is inherent to what we identify as the in situ component.\par
We posit that the \(\chemZH-\sigma_z\) relation is driven by the successive `generations' of star formation, each becoming more enriched and more dynamically-cold than those before (in the absence of accretion which would result in the chemical and dynamical mixing of the populations). This could be the case if, for instance, mass segregation of metals occurs vertically as well as radially. Alternatively, this would be the result if higher-metallicity gas requires colder kinematics before star formation is possible, or if the cooling effects of metals naturally produces more dynamically-cold disks if the gas is more metal-rich. \cite{choi2009} show that metal cooling can significantly increase the star-formation efficiency of the inter-stellar medium, though there is no direct link in that work to dynamics.\par
Yet measurements of the gas-phase AVR show clear trends with redshift, and the physical interpretation in that case explicitly includes a redshift dependence. However this redshift dependence is via gas depletion through the cosmic specific SFR \citep[such as in][]{whitaker2014}; in that scenario, galaxies with larger gas fractions experience larger inter-stellar medium turbulence, and higher \(\sigma_z\) is imparted to the stars upon star formation \citep{leaman2017}. So in much the same way as the scenario proposed here for the stellar AVR, the gas-phase AVR is tied to episodes of star formation, which happen to decline on average with redshift. This subtle difference is especially important when analysing individual galaxies with individual assembly histories. There also remains significant scatter at fixed redshift within the gas-phase AVR that needs to be accounted for, which indicates a potential additional dimension to this issue. Naturally, these star-formation episodes lead to enrichment of the gas over cosmic time \citep{daigne2006, kobayashi2007}. Therefore, if metallicity is the underlying physical driver of the gas-phase AVR, it would still manifest as an observed redshift dependence without intrinsically depending on redshift directly. But this is only, at present, a circumstantial argument in lieu of an explicit experiment for gas disks.\par
In any case, a testable prediction of this hypothesis is that at fixed present-day stellar mass (and without significant ex situ contributions or perturbations), galaxies with higher SFR (that is, faster chemical enrichment) should achieve dynamically-colder orbits at fixed age --- or alternatively, the stellar AVR should have a steeper slope. This is because in such a scenario, the absolute cosmic time is not the driver of the AVR, but rather the time it takes for a particular galaxy to achieve a particular degree of enrichment. Since the stellar metallicity and \(\sigma_z\) will depend on both stellar mass and accretion history, it is imperative to control for those parameters to test this prediction. This is at present not possible for the sample of galaxies for which our analysis has been performed, but should be accessible to theoretical models and simulations. In fact, \cite{just2010a} explicitly investigate the effect of SFR on the stellar AVR, tailored to fit the Milky-Way, through a series of models. That work finds that for similar forms of the SFH, the model with a higher SFR has lower vertical velocity dispersion, despite peaking at earlier epochs.\par
The outlier in this respect from \cref{img:fixedSPDisp} is NGC~3115. We have already established that accretion has played a minor role in the stellar mass assembly of the three Fornax galaxies. Conversely, \cite{poci2019} conclude that NGC~3115 assembled \(\sim 68\%\) of its present-day stellar mass from external sources. Moreover, given the higher stellar mass of NGC~3115, these accreted systems could be higher mass, and therefore more enriched on average, compared to lower-mass satellites accreted onto lower-mass hosts. Thus the age and metallicity trends would be significantly phase-mixed, as is seen by the reduced correlation coefficients. The persistence of an AVR, only at high metallicity, may be indicative of secular evolution following the last accretion event.\par
A similar analysis has been performed for the Milky-Way using a combination of many of the recent photometric and spectroscopic surveys \citep{sharma2020}. That work finds a strong \(\chemZH-\sigma_z\) relation at fixed age. Yet they also find a persistent stellar AVR at fixed metallicity, but only at young ages. The AVR then flattens and correlates solely with metallicity at old ages. \cite{yu2018} see similar trends with two bins of metallicity. \cite{sharma2020} interpret the \(\chemZH-\sigma_z\) correlation as a connection between \(\sigma_z\) and the stellar birth radius. However, in the case of the Fornax galaxies, we see no clear (monotonic) radial gradients of metallicity in \cref{img:2dmap153,img:2dmap170,img:2dmap177}. Comparisons to that work are complicated, however, by the selection functions of the Milky-Way data sets, and so these results may be tracing different physical regimes.
\subsection{\SZ\ formation}\label{ssec:sz}
All of our results indicate that the Fornax galaxies have undergone mild transformations due to the cluster potential, primarily in their outer regions. They have been able to retain their cold central kinematics, yet in a thicker configuration compared to the field. We posit, thus, that their \SZ\ morphology is a result of these interactions. This is neither of the explanations typically invoked to explain the transformations of galaxies into \SZ; mergers \citep{chilingarian2009, querejeta2015, tapia2017, poci2019} or the `fading' of spiral galaxies \citep{larson1980, bekki2002, donofrio2015, mishra2018, rizzo2018}. While cluster environments are common in the faded-spiral scenario, it is predicated on the supposed spiral progenitor first ceasing star-formation due to the environment \citep[for instance,][]{boselli2006, book2010, peng2012, mendel2013, bekki2014}, allowing it to subsequently transform into an \SZ\ galaxy. Yet we see evidence for significant star formation activity well beyond the suspected time of in-fall to the cluster for all three galaxies. Gas was therefore readily available until relatively recently. The evidence for a lack of stellar accretion has been discussed, in agreement with other cluster studies, rendering this formation path unlikely as well. This is also consistent with the deductions of \cite{comeron2019} and \cite{pinna2019a} who find that accretion does not play a major role in the formation of `thick' disks in local galaxies.\par
An alternative scenario proposed by \cite{diaz2018} states that at high redshift, gas-rich satellite accretion onto compact elliptical galaxies leads to the formation of the disk component of the resulting \SZ. Our models impose the constraint that if this scenario occurred for the three Fornax galaxies, such accretion would have had to occur \(\gtrsim 12\ \si{\giga\year}\) ago for FCC~153 and FCC~170, and \(\gtrsim 10\ \si{\giga\year}\) ago for FCC~177, since it must precede the formation of the dynamically-cold disk. However, this scenario supposes that the compact elliptical, which goes on to form the `bulge' of the subsequent \SZ, is responsible for the suppression of spiral arms in the disk. Conversely, our data suggest that only FCC~170 has a significant contribution from a central pressure-supported structure. More broadly, a diversity of \SZ\ properties is emerging \citep{fraser-mckelvie2018, coccato2020, deeley2020, tous2020}, and it is unlikely that a single formation path is responsible for this diversity.
\subsection{The Fornax galaxy cluster population}
The photometric catalogue of \cite{ferguson1989}, covering \(40\) sq. degrees, contains \(35\) \SZ-like galaxies (some of which have uncertain classification), with \(20\) being brighter than the magnitude limit of the \ftd\ survey \((m_B \leq 15)\). Of these, \(12\) were observed by \ftd, accounting for \(60\%\ (34\%)\) of bright (all) lenticular galaxies in the Fornax cluster. Our analysis on the sub-sample of three galaxies, chosen for the reasons discussed above, can not therefore account for the expected diversity of evolutionary pathways within the cluster's galaxy population. So while we infer no major mergers for these galaxies, for instance, we can not exclude this formation path for some of the other cluster \SZ\ galaxies. We have, however, probed the relative extremes in terms of assembly histories, with FCC~170 and FCC~177 forming the majority of their stellar content at early and late times, respectively. This analysis has simultaneously uncovered properties which appear to be approximately independent of assembly history - namely the stellar AVR - which we thus expect to hold for all but the most violent histories. To confidently infer the histories of the remaining galaxies from \ftd, this analysis must be applied to each of them individually. This is the goal of future work.

\section{Conclusions}
In this work we modelled three edge-on \SZ\ galaxies in the Fornax cluster as part of the \ftd\ project. We applied sophisticated dynamical and stellar-population techniques to self-consistently model the entire stellar information content. These models were used to infer how each galaxy formed, and allowed us to place strong constraints on some of the hypothesised processes that affect galaxy formation and evolution, in particular in the cluster environment. These findings are summarised here:
\begin{itemize}[label=\(\bullet\)]
    \item All three galaxies retain a strongly-rotating component that has persisted for many dynamical times. These structures can be composed of both young and old stars, implying that they have survived the galaxy's entry into the cluster and subsequent evolution therein (\cref{img:mfh153,img:mfh170,img:mfh177}).
    \item There is evidence of continued star-formation in all three galaxies, to varying degrees. Owing to the metallicity and kinematics, we suggest this star formation is almost exclusively in-situ through recycled gas, as there is no evidence of gas accretion (\cref{img:mfh153,img:mfh170,img:mfh177}).
    \item Our results are suggestive of a suppression of stellar accretion. We postulate that this is driven by the relative motions of galaxies within the cluster (\cref{img:mfh153,img:mfh170,img:mfh177}), as proposed in previous works.
    \item There is evidence against internal heating as the cause of the stellar age--velocity-dispersion relation, suggesting that older stars were created with inherently-higher \(\sigma_z\), in agreement with the results of \cite{poci2019}. There is tentative evidence that the relations for the cluster galaxies are elevated with respect to the field (\cref{img:cosmoDisp}), implying that harassment may be responsible for mild dynamical heating.
    \item We find tentative observational evidence of a potential fundamental stellar \(\chemZH-\sigma_z\) relation which we argue may contribute significantly to the observed age--velocity-dispersion relation (\cref{img:fixedSPDisp}).
\end{itemize}\par
We endeavour in future work to incorporate more detailed stellar-population analyses, including variable IMF, while continuing to apply this methodology to a variety of galaxies. Deriving these histories for other galaxies will enable a more thorough understanding of how galaxies piece together their mass, and which processes have dominant effects in various regimes.


\begin{acknowledgements}
We thank Lorenzo Morelli, Thomas Spriggs, and Adrian Bittner for discussions on this work. AP acknowledges financial support from Macquarie University and the ESO Studentship Programme. RMM is the recipient of an Australian Research Council Future Fellowship (project number FT150100333). LZ acknowledges the support from National Natural Science Foundation of China under grant No. Y945271001, and the National Key R$\&$D Program of China under grant No. 2018YFA0404501. GvdV acknowledges funding from the European Research Council (ERC) under the European Union's Horizon 2020 research and innovation programme under grant agreement no. 724857 (Consolidator Grant ArcheoDyn). EMC is supported by MIUR grant PRIN 2017 20173ML3WW{\_}001 and by Padua University grants DOR1715817/17, DOR1885254/18, and DOR1935272/19. JF-B, IMN, and FP acknowledge support through the RAVET project by the grant PID2019-107427GB-C32 from The Spanish Ministry of Science and Innovation.
\par
Based on observations collected at the European Southern Observatory under ESO programme 296.B-5054(A). This work makes use of the \tfo{SciGar} compute cluster at ESO, and the \tfo{OzStar} supercomputer at Swinbourne University. The work also makes use of existing software packages for data analysis and presentation, including \tso{AstroPy} \citep{astropycollaboration2013}, \tso{Cython} \citep{behnel2011}, \tso{IPython} \citep{perez2007}, \tso{matplotlib} \citep{hunter2007}, \tso{NumPy} \citep{harris2020a}, the \tso{SciPy} ecosystem \citep{virtanen2020}, and \tso{statsmodels} \citep{seabold2010}. We finally thank the anonymous referee, whose feedback greatly improved the depth and clarity of this work.
\end{acknowledgements}


\bibliographystyle{aa}
\bibliography{f3dIv8}


\begin{appendix}
\section{Mass density MGE}\label{app:massMGE}

\begin{figure}
    \centerline{
        \includegraphics[width=\columnwidth]{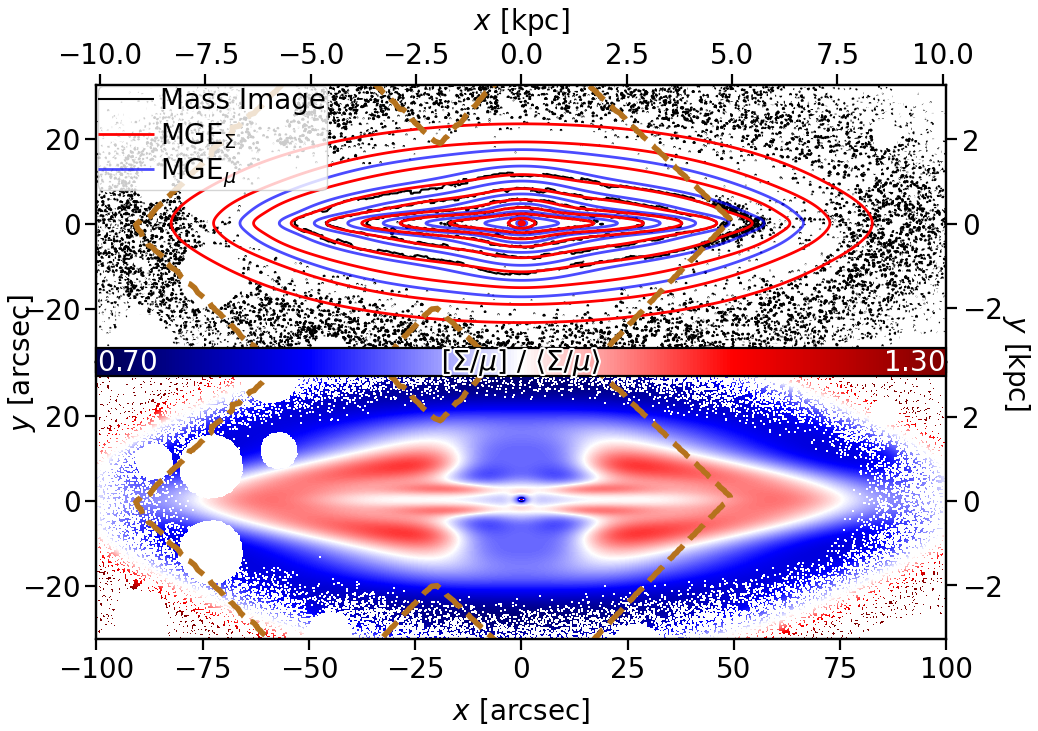}
    }
    \caption{the mass-density MGE for FCC~153. {\em Top:} Contours of the projected mass density (black), the mass density \mgeT\ (red), and the surface brightness \mgeS\ (blue). {\em Bottom:} The ratio between the mass and surface-brightness models, showing the structural differences. This panel is normalised by the average \(M_\star/L\) such that values of \(1.0\) (white) are consistent with the assumption of a spatially-constant \(M_\star/L\), and any deviations are percentages in stellar mass. The outline of the MUSE mosaic is shown in dashed brown.}
    \label{img:153massMGE}
\end{figure}

\begin{figure}
    \centerline{
        \includegraphics[width=\columnwidth]{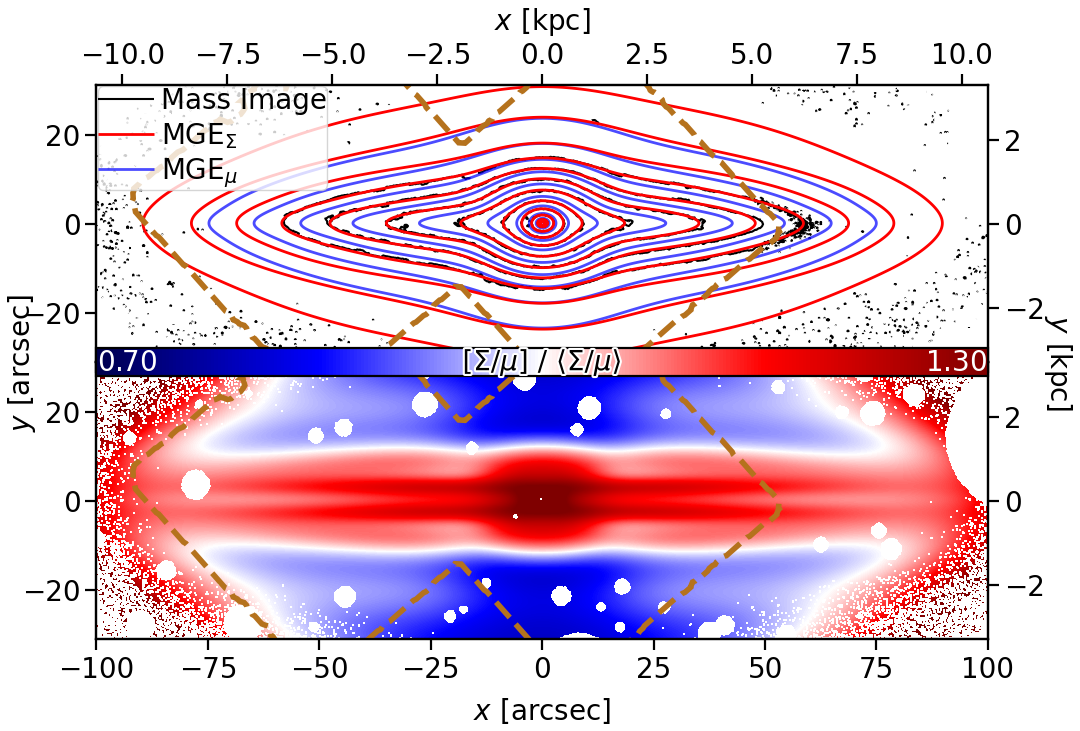}
    }
    \caption{Same as \protect\cref{img:153massMGE}, but for FCC~170.}
    \label{img:170massMGE}
\end{figure}

\begin{figure}
    \centerline{
        \includegraphics[width=\columnwidth]{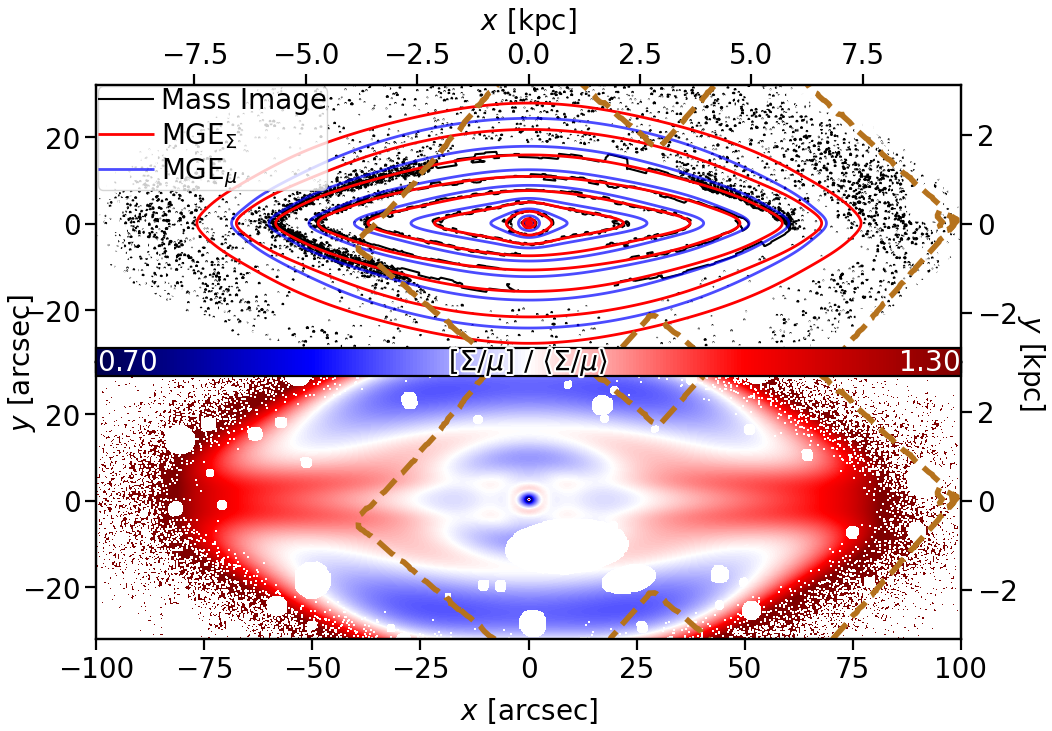}
    }
    \caption{Same as \protect\cref{img:153massMGE}, but for FCC~177.}
    \label{img:177massMGE}
\end{figure}

Here we present the fits to the scaled mass `images' described in \cref{ssec:massModel}, in \cref{img:153massMGE,img:170massMGE,img:177massMGE}, and the results themselves in \cref{tab:153massMGE,tab:170massMGE,tab:177massMGE}, for FCC~153, FCC~170, and FCC~177, respectively. The MGE in this work have their major-axis offsets \(\psi\) fixed to zero, which generates axisymmetric (in projection) mass models.
\begin{table}
	\begin{tabular}{S[table-format=6.2]|S[table-format=3.3]|c}
	 $\Sigma$ & $\sigma$ & $q$\\
	 $[\si{\Msun\per\parsec\squared}]$ & $[{\rm arcsec}]$ & \\
	\hline
	68049.68 & 0.278 & 0.52524\\
	4478.66 & 1.439 & 0.59253\\
	2013.16 & 4.619 & 0.51940\\
	357.27 & 6.725 & 0.81297\\
	904.26 & 14.553 & 0.15344\\
	4127.99 & 18.123 & 0.06621\\
	582.02 & 26.947 & 0.15240\\
	300.95 & 31.695 & 0.29016\\
	12.17 & 74.105 & 0.53099\\
	\hline
	\end{tabular}
    \caption{\protect\mgeT\ for FCC~153. The columns represent, from left to right, the projected mass surface density, the width (peak location), and axis ratio, respectively.}
    \label{tab:153massMGE}
\end{table}
\begin{table}
	\begin{tabular}{S[table-format=6.2]|S[table-format=3.3]|c}
	 $\Sigma$ & $\sigma$ & $q$\\
	 $[\si{\Msun\per\parsec\squared}]$ & $[{\rm arcsec}]$ & \\
	\hline
	177222.31 & 0.358 & 0.73304\\
	44814.95 & 1.198 & 0.74880\\
	17514.47 & 2.819 & 0.75675\\
	5956.28 & 6.776 & 0.76166\\
	127.74 & 12.580 & 0.99900\\
	2229.11 & 23.103 & 0.10200\\
	1113.45 & 29.559 & 0.18741\\
	170.86 & 38.541 & 0.33380\\
	18.98 & 68.965 & 0.71089\\
	3.06 & 88.782 & 0.20664\\
	1.81 & 88.782 & 0.99900\\
	\hline
	\end{tabular}
    \caption{Same as \cref{tab:153massMGE}, but for FCC~170.}
    \label{tab:170massMGE}
\end{table}
\begin{table}
	\begin{tabular}{S[table-format=6.2]|S[table-format=3.3]|c}
	 $\Sigma$ & $\sigma$ & $q$\\
	 $[\si{\Msun\per\parsec\squared}]$ & $[{\rm arcsec}]$ & \\
	\hline
	61581.78 & 0.231 & 0.79514\\
	1601.36 & 1.497 & 0.80274\\
	915.55 & 3.190 & 0.72995\\
	315.87 & 9.467 & 0.45654\\
	13.62 & 20.762 & 0.99701\\
	363.99 & 22.622 & 0.07590\\
	463.00 & 25.577 & 0.17350\\
	174.70 & 29.337 & 0.35693\\
	8.21 & 50.489 & 0.68817\\
	7.51 & 64.291 & 0.23789\\
	\hline
	\end{tabular}
    \caption{Same as \cref{tab:153massMGE}, but for FCC~177.}
    \label{tab:177massMGE}
\end{table}
This section illustrates a number of key aspects of this process. Firstly, it can be seen from the upper panels of \cref{img:153massMGE,img:170massMGE,img:177massMGE} that the transition from spectroscopically-derived to photometric-derived \(M_\star/L_R\) is seamless, albeit with higher noise in the photometry. Conversely, systematic over- and under-estimations of \(M_\star/L_R\) from the photometry would present as discontinuous shifts to larger and smaller radii, respectively, of a given contour at the transition between photometry and spectroscopy. Since there is no such discontinuity in these results, it indicates that the photometrically-derived \(M_\star/L_R\) values are quantitatively consistent with those derived from spectroscopy. Secondly, the lower panels clearly highlight the shape differences between the luminosity and mass surface densities. This is a direct result of the resolved structures in stellar populations, which are subsequently taken into account in the dynamical models by this approach.

\section{Dynamical modelling}\label{app:schwarz}
\begin{figure}
    \centerline{
        \includegraphics[width=\columnwidth]{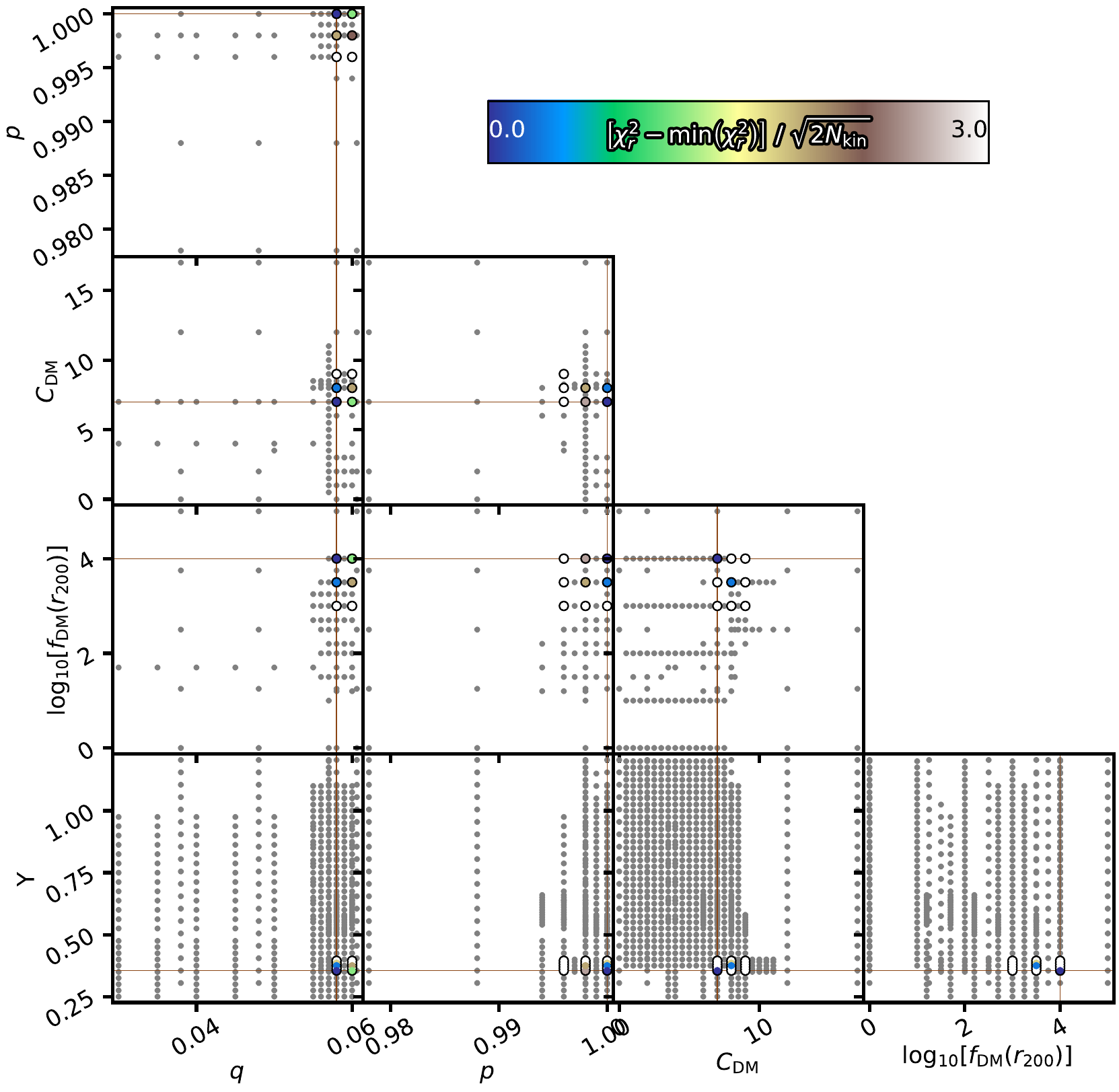}
    }
    \caption{Optimisation over the \ND{5} parameter-space for FCC~153. The grey points indicate the exploration of the parameter-space with the smaller orbit sampling, while the coloured points show the \(\chi_r^2\) of the models with the high orbit sampling (see \cref{ssec:schwarz}). The best-fit parameters are indicated by the brown lines.}
    \label{img:corn153}
\end{figure}

\begin{figure}
    \centerline{
        \includegraphics[width=\columnwidth]{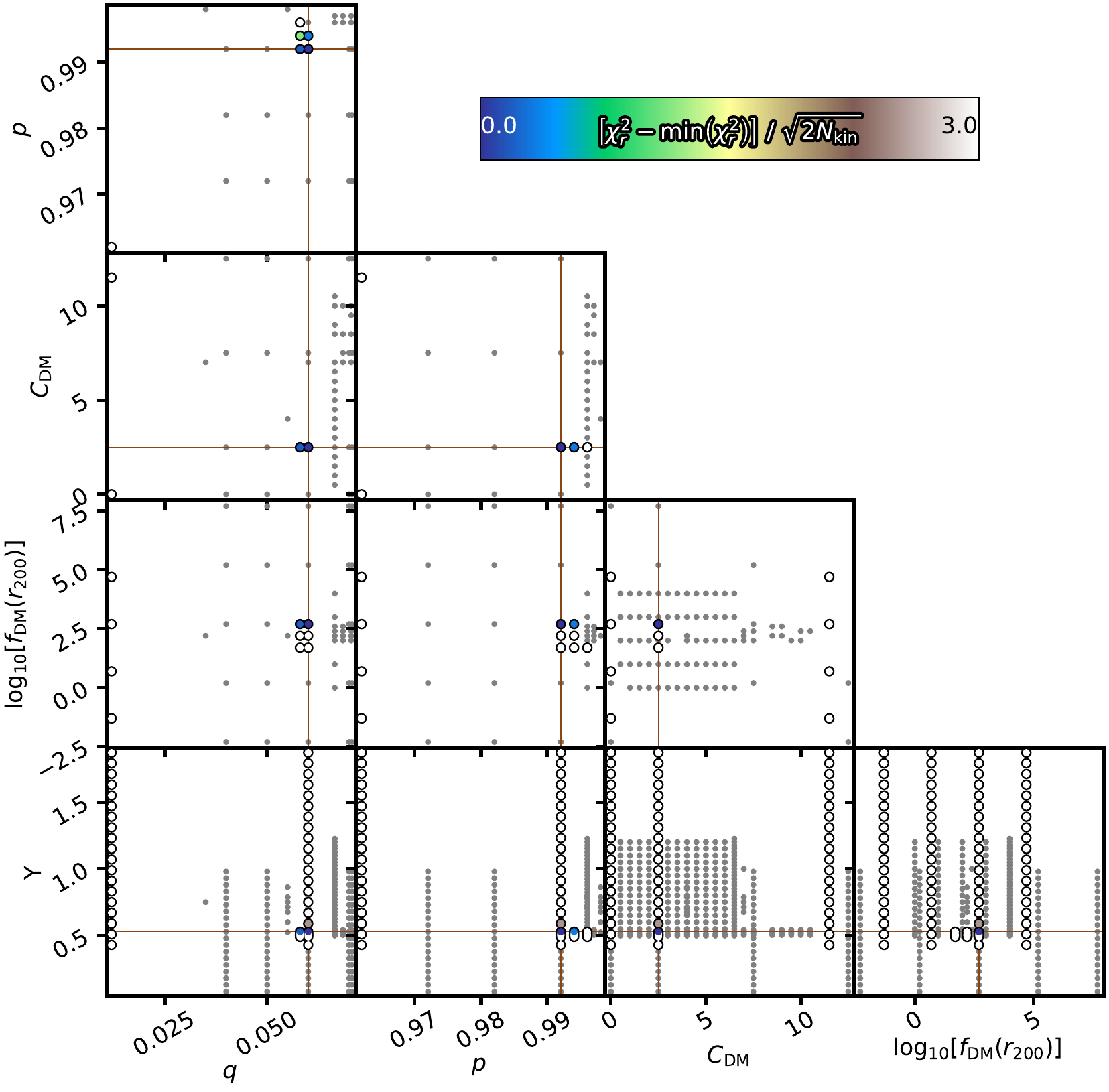}
    }
    \caption{Same as \protect\cref{img:corn153}, but for FCC~170.}
    \label{img:corn170}
\end{figure}

\begin{figure}
    \centerline{
        \includegraphics[width=\columnwidth]{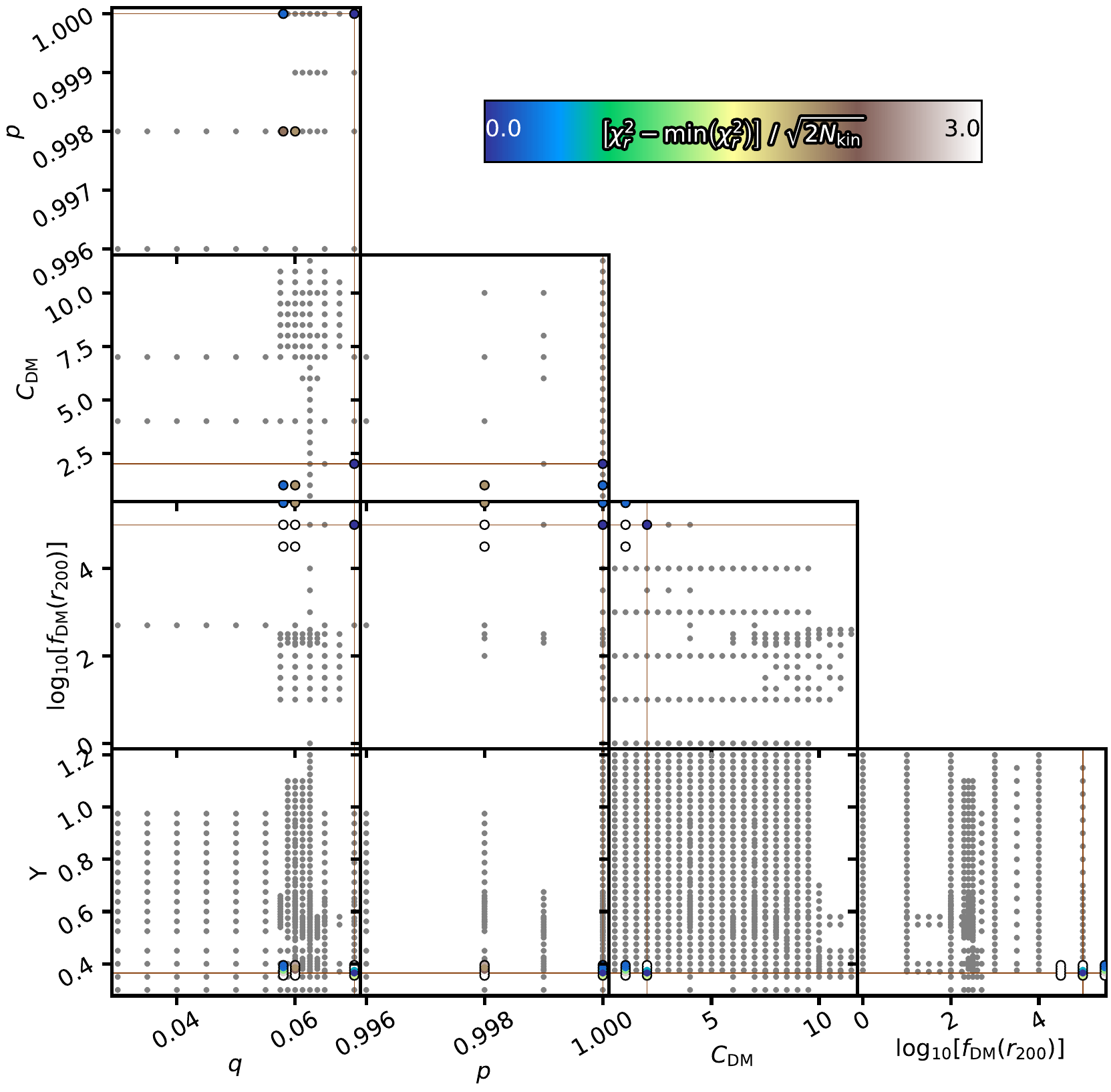}
    }
    \caption{Same as \protect\cref{img:corn153}, but for FCC~177.}
    \label{img:corn177}
\end{figure}

\begin{figure}
    \centerline{
        \includegraphics[width=\columnwidth]{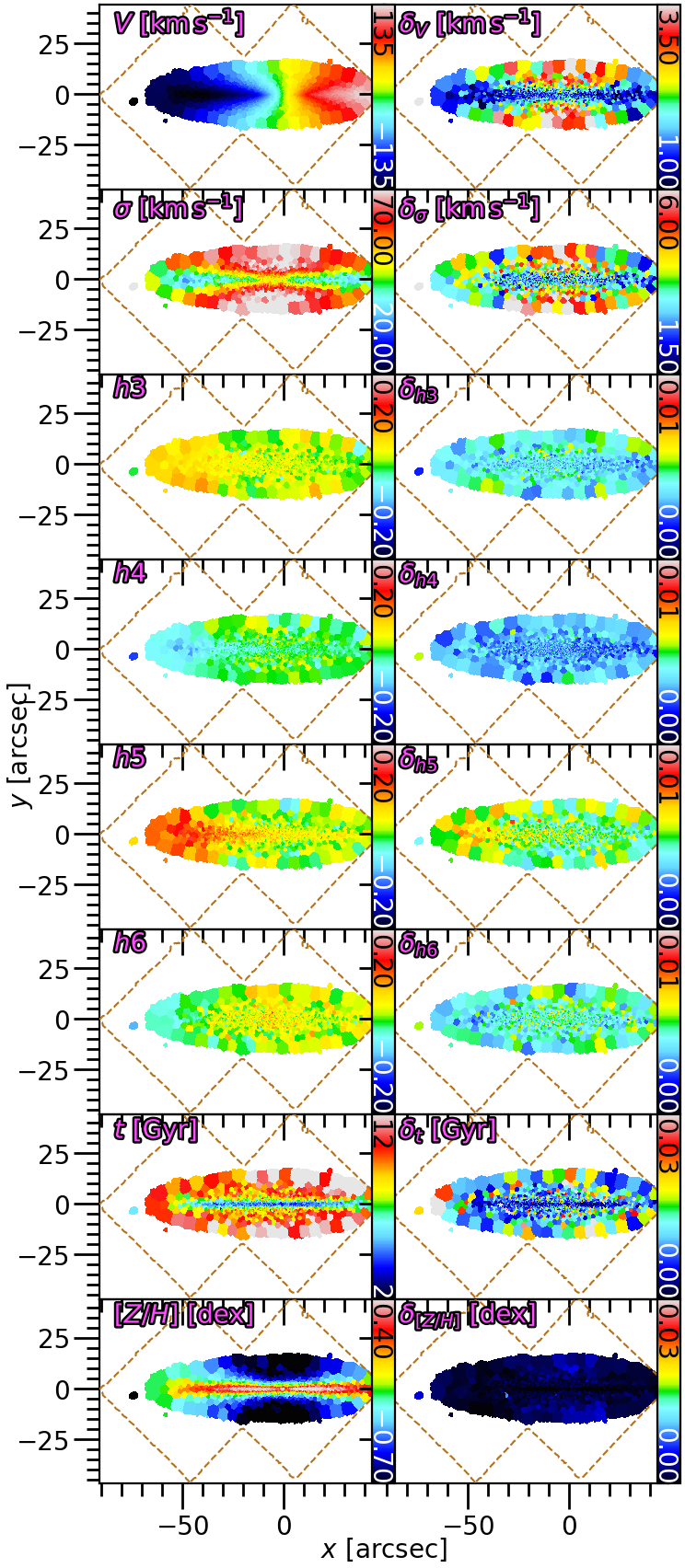}
    }
    \caption{Un-symmetrised kinematics extracted for FCC~153 ({\em left}), and associated errors computed through Monte Carlo simulations ({\em right}). The outline of the MUSE mosaic is shown in dashed brown.}
    \label{img:raw153}
\end{figure}

\begin{figure}
    \centerline{
        \includegraphics[width=\columnwidth]{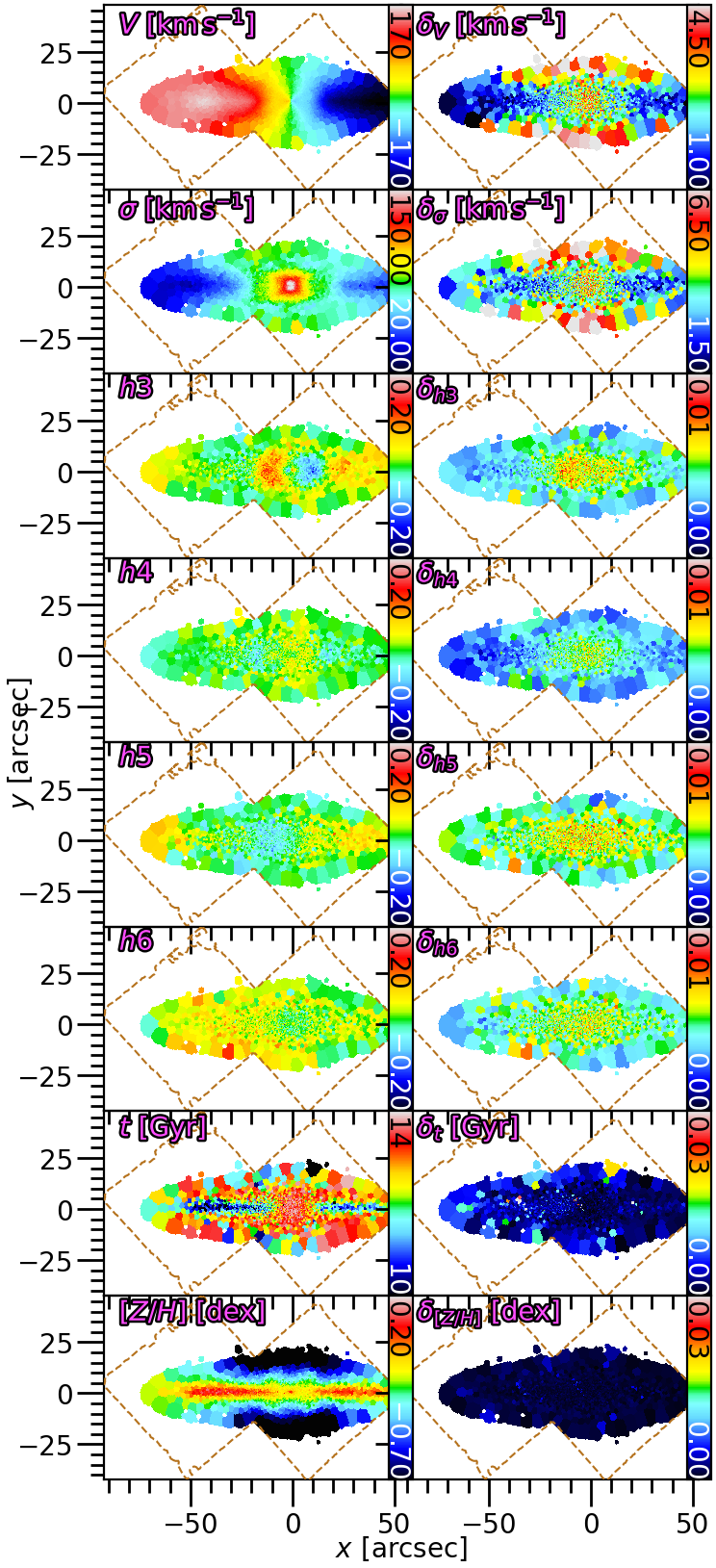}
    }
    \caption{Same as \protect\cref{img:raw153}, but for FCC~170.}
    \label{img:raw170}
\end{figure}

\begin{figure}
    \centerline{
        \includegraphics[width=\columnwidth]{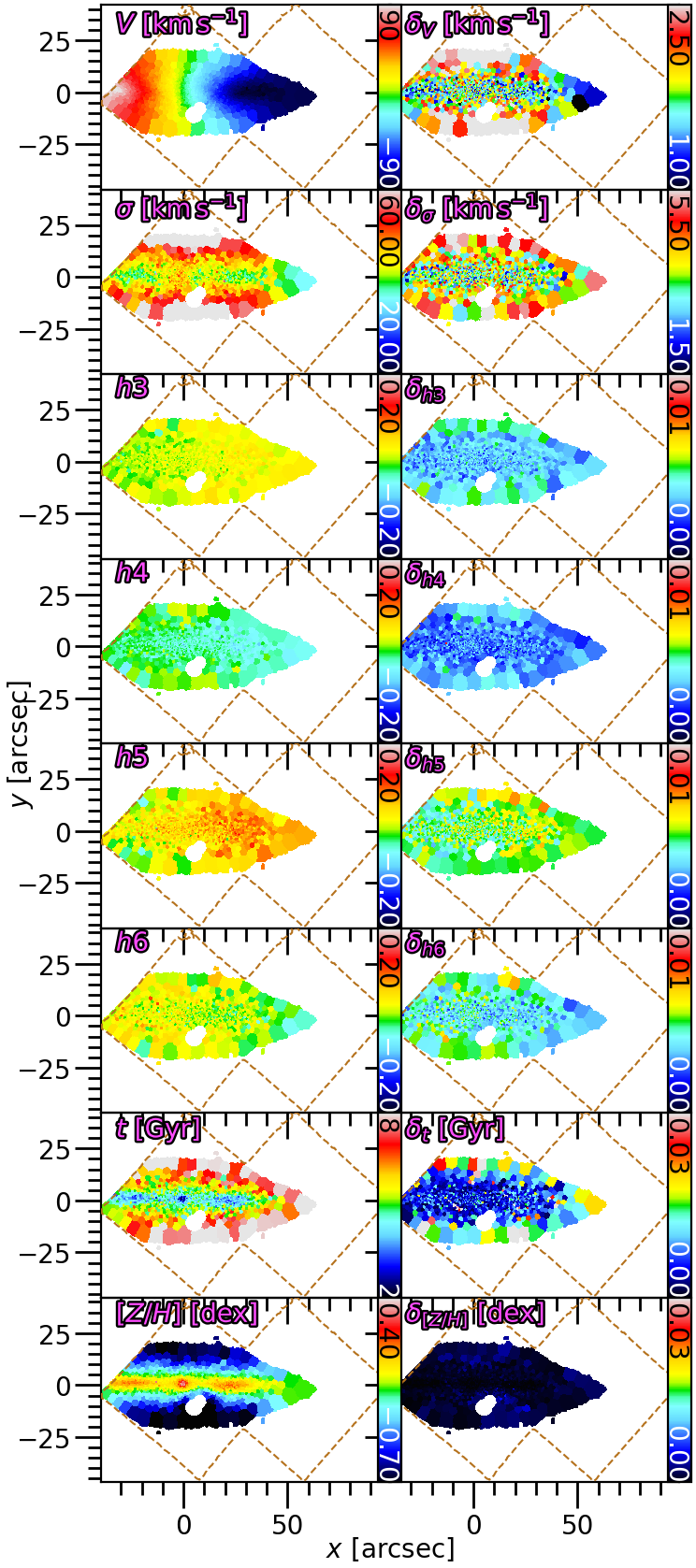}
    }
    \caption{Same as \protect\cref{img:raw153}, but for FCC~177.}
    \label{img:raw177}
\end{figure}

This section presents additional details of the \shw\ models and measured data products. The best-fit parameters of the \shw\ models are given in \cref{tab:schwarz}, while their parameter-space distributions are shown in \cref{img:corn153,img:corn170,img:corn177}. The measured (un-symmetrised) kinematics and stellar populations are presented in \cref{img:raw153,img:raw170,img:raw177}.
\begin{table*}
    \(\begin{tabu}{cccccccc}
        {\rm Galaxy} & \log_{10}(M_\bullet\ [\si{\Msun}])^* & q & p & u^* & C_{\rm DM} & \log_{10}(f_{\rm DM}) & \Upsilon\\\hline
        {\rm FCC~153} & 6.03 & 0.0580 & 0.9980 & 0.9999 & 7.00 & 4.00 & 0.355\\
        {\rm FCC~170} & 7.40 & 0.0600 & 0.9920 & 0.9999 & 2.50 & 2.70 & 0.530\\
        {\rm FCC~177} & 5.52 & 0.0700 & 0.9999 & 0.9999 & 2.00 & 5.00 & 0.365\\\hline
    \end{tabu}\)
    \caption{Best-fit parameters for the \shw\ models. Columns marked with \({}^*\) are fixed to the values given here.}
    \label{tab:schwarz}
\end{table*}
It can be seen that FCC~170 exhibits the highest degree of intrinsic triaxiality, with the lowest \(p\), though the triaxiality is small in all cases. The intrinsic shapes recovered by our models are in good agreement with, for instance, those of the oblate galaxies of \cite{jin2019}.  Interestingly, the dynamical model of the field \SZ\ NGC~3115 \citep{poci2019} has larger intrinsic triaxiality than the three cluster \SZ\ galaxies studied here, but this is likely a consequence of its higher stellar mass, and the more violent assembly history inferred for that galaxy.

\section{Validation on mock data}\label{sec:mockvalid}
A crucial test to conduct for the modelling procedure used in this work is to determine the accuracy and precision with which it can recover known properties. To this end, we conducted tests on mock data of realistic galaxies from the Auriga simulation suite \citep{grand2017}. In \cite{zhu2020}, this \shw\ code is used to fit the mock kinematic data of sub-halos from Auriga, at various projected inclinations. Here we take the dynamical models and apply the procedure described in \cref{ssec:ddec,ssec:dynPop} in order to fit stellar populations in exactly the manner that is applied to the \ftd\ data. We then measure the diagnostics that inform the main scientific conclusions of this work, and compare directly to the underlying intrinsic distributions from the simulations. We conduct this analysis on Auriga halos 5 and 6, projected to an inclination of \(\theta=80\si{\degree}\)\par
\begin{figure}
    \centerline{
        \includegraphics[width=0.8\columnwidth]{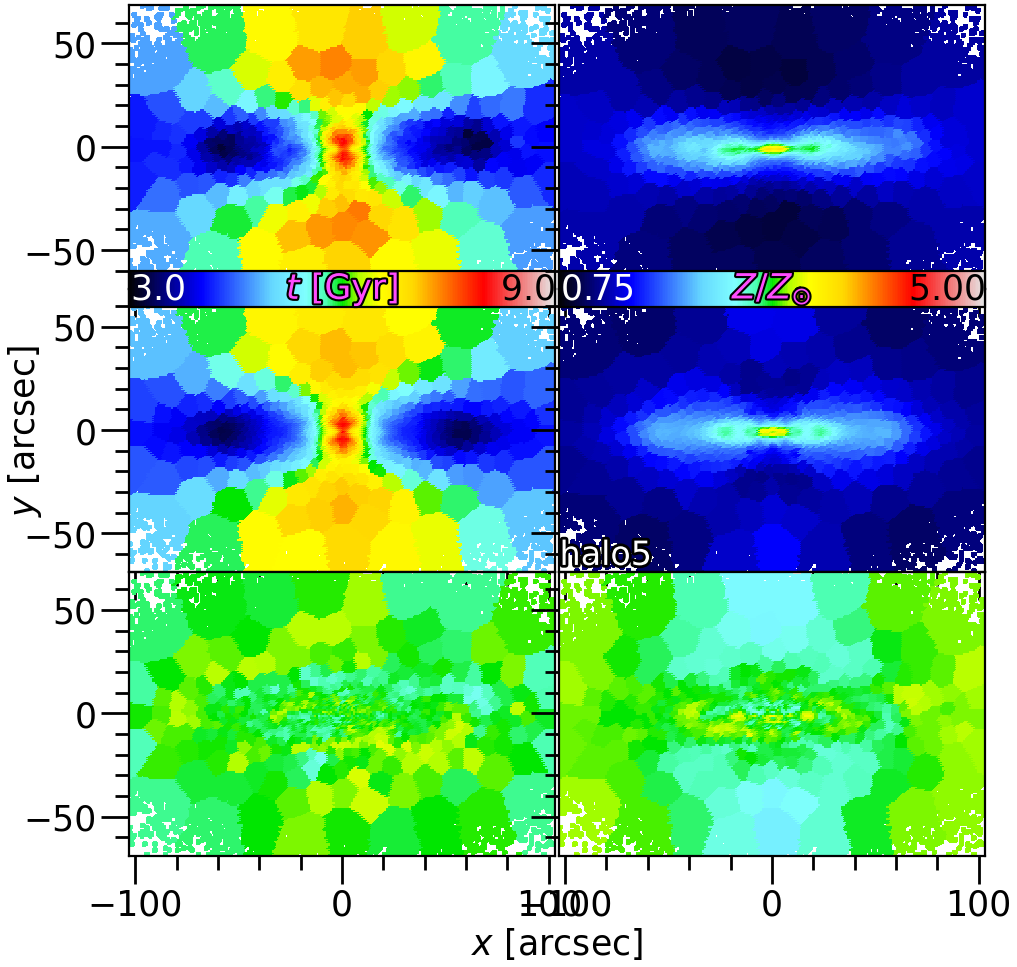}}
    \centerline{
        \includegraphics[width=0.8\columnwidth]{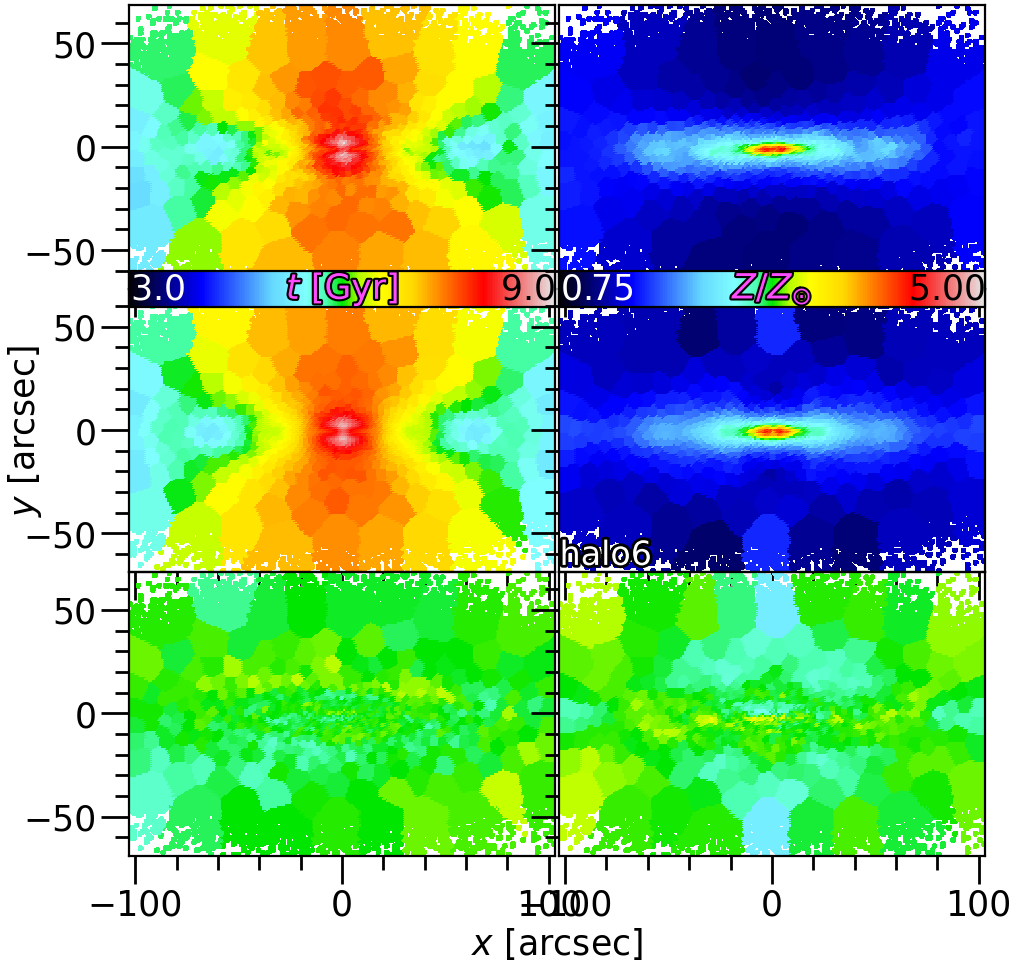}
    }
    \caption{Stellar-population \shw\ model fits to the mock data. Mean stellar age ({\em left}) and metallicity ({\em right}) are shown for Auriga halos 5 ({\em top}) and 6 ({\em bottom}). From top to bottom, the rows represent the true means, models fits, and residuals.}
    \label{img:mockMW}
\end{figure}
\begin{figure}
    \centerline{\includegraphics[width=\columnwidth]{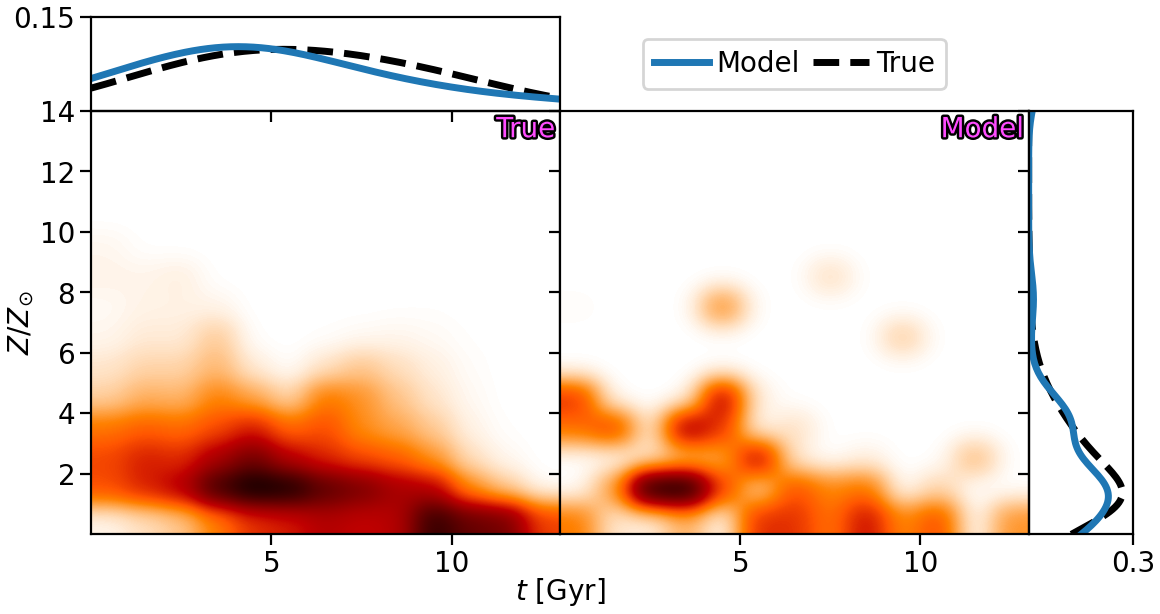}}
    \centerline{\includegraphics[width=\columnwidth]{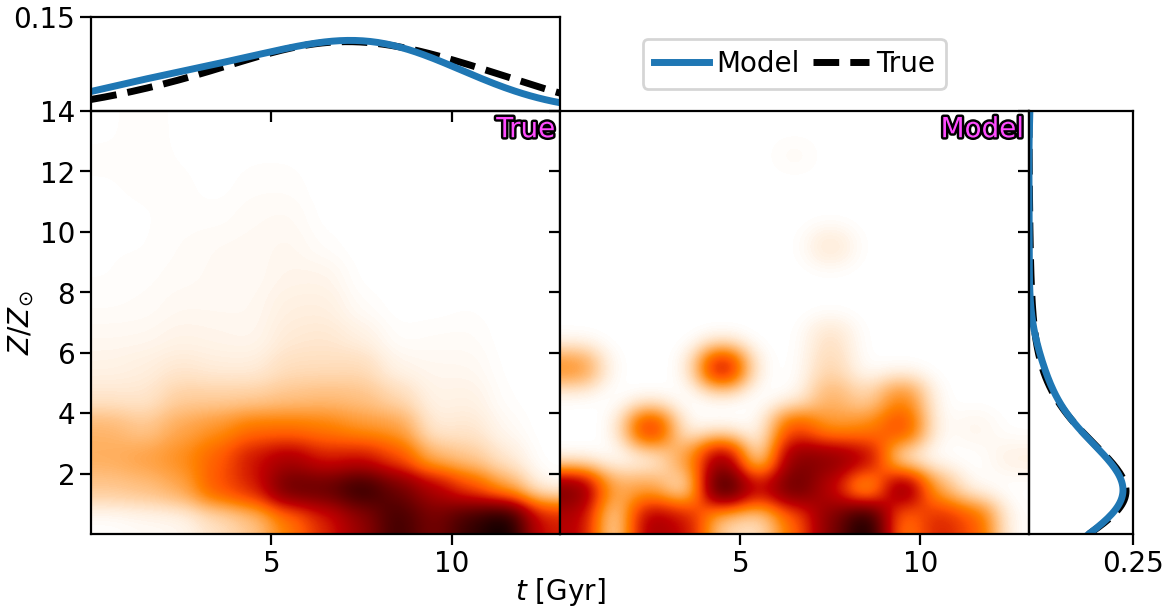}}
    \caption{Intrinsic \(t-Z\) distribution ({\em left}) and that retrieved from our model ({\em right}) for halos 5 ({\em top}) and 6 ({\em bottom}). The \ND{1} distributions are also shown for each simulation, as kernel density estimates of the underlying data. The intrinsic distributions are black dashed curves, and the model distributions are the blue solid curves.}
    \label{img:mockSFH}
\end{figure}

The stellar-population fits are shown in \cref{img:mockMW}, where the `true' maps are generated simply by projecting all the particles along each LOS. We also ran \(100\) Monte Carlo simulation fits to these maps as is done in the main text, however in this case the outputs are less informative because the `errors' on the mock maps were generated artificially. The distributions of \(t-Z\) are presented in \cref{img:mockSFH}. We note here that the `true' distributions are smoother due to the \(\sim 10^5\) particles that are used to generate them, in contrast to the \(\sim 10^2\) dynamical components which constitute the `model' distributions. Kernel Density Estimates (KDE) were computed from these data to compare the marginalised distributions. Owing to the difference in the size of the data sets between the dynamical models and simulations, we derived the optimal bandwidth for the KDE of each data-set using a data-driven approach which minimises the variance and also takes into account each sample size \citep{silverman1986}. This exercise has shown that not only can the models fit the projected maps, but also that our implementation recovers the underlying stellar-population distributions.\par
The main results from this work are derived from the assembly histories in \cref{img:mfh153,img:mfh170,img:mfh177}, and the AVR in \cref{img:cosmoDisp}. Comparing simulations to data in general can introduce inconsistencies in many ways, and so for clarity we describe the procedure for testing the recovery of the assembly histories on simulations, and the steps taken to mitigate potential inconsistencies. Firstly, the Cartesian coordinates of the simulation data are binned to the same spherical polar grid of the corresponding \shw\ model. The simulation kinematics, \(V_x, V_y, V_z\), are then converted to cylindrical coordinates, in line with how the outputs from the \shw\ model are analysed. From here, the \(\sigma_z\) measurements on the simulation data proceed in an identical fashion as for the \shw\ model. Moreover, the distributions from the simulations are constructed using exactly the particles that lie within the mock FOV, by construction. Conversely, the \shw\ models (and indeed real observations) typically have contributions from orbits which reside (on average) outside the FOV, which are not present in the simulation. This means that any measurement on these data would be probing different physical regions between the two data sets. To remedy this inconsistency, the results from the \shw\ models do not include orbits with time-averaged radii outside the FOV. This is only the case for the mock tests where strictly-consistent comparisons are sought. Our main results in \cref{img:mfh153,img:mfh170,img:mfh177} include the full model.\par
\begin{figure*}
    \centerline{
        \includegraphics[width=\textwidth]{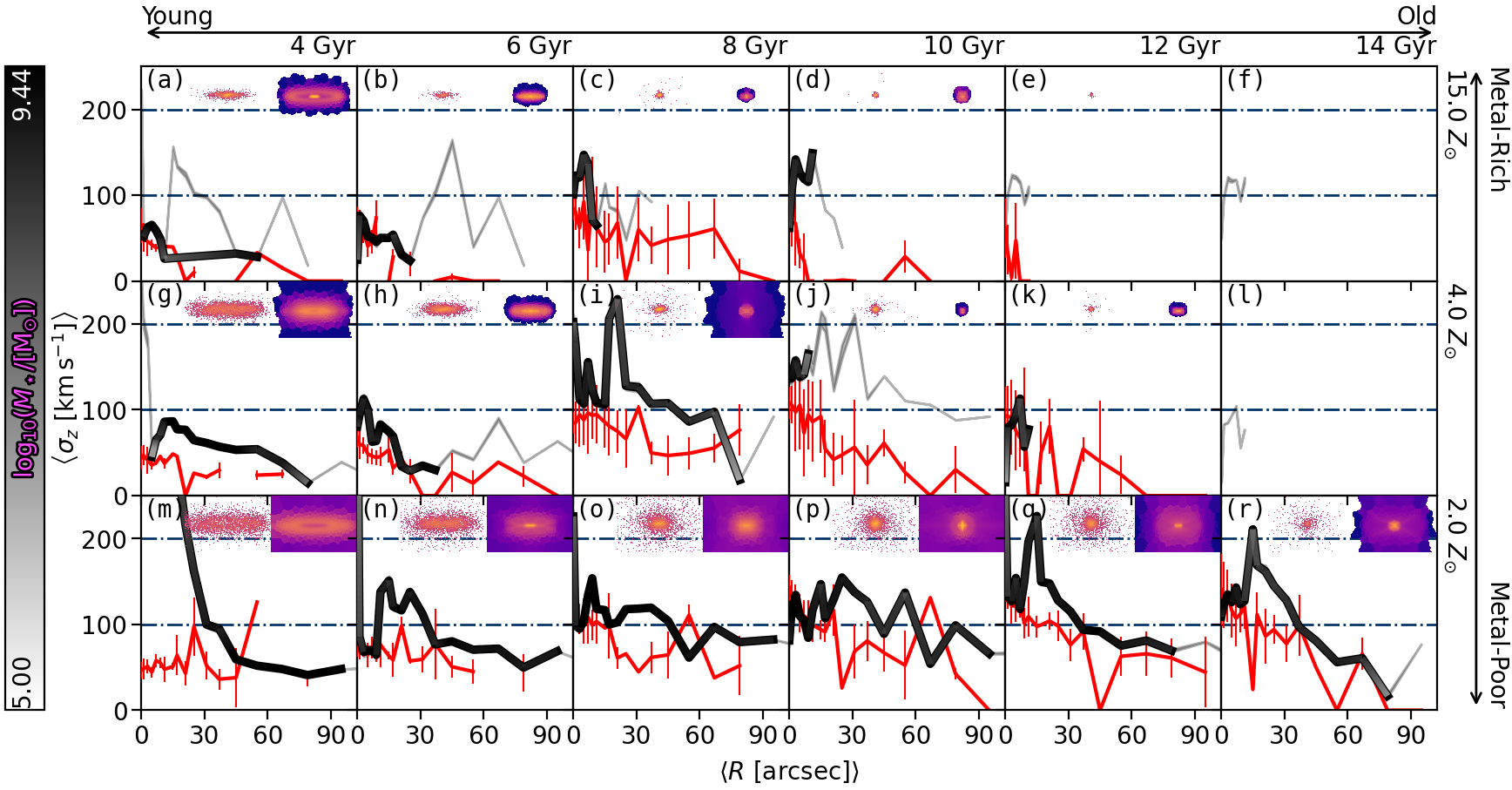}}
    \centerline{
        \includegraphics[width=\textwidth]{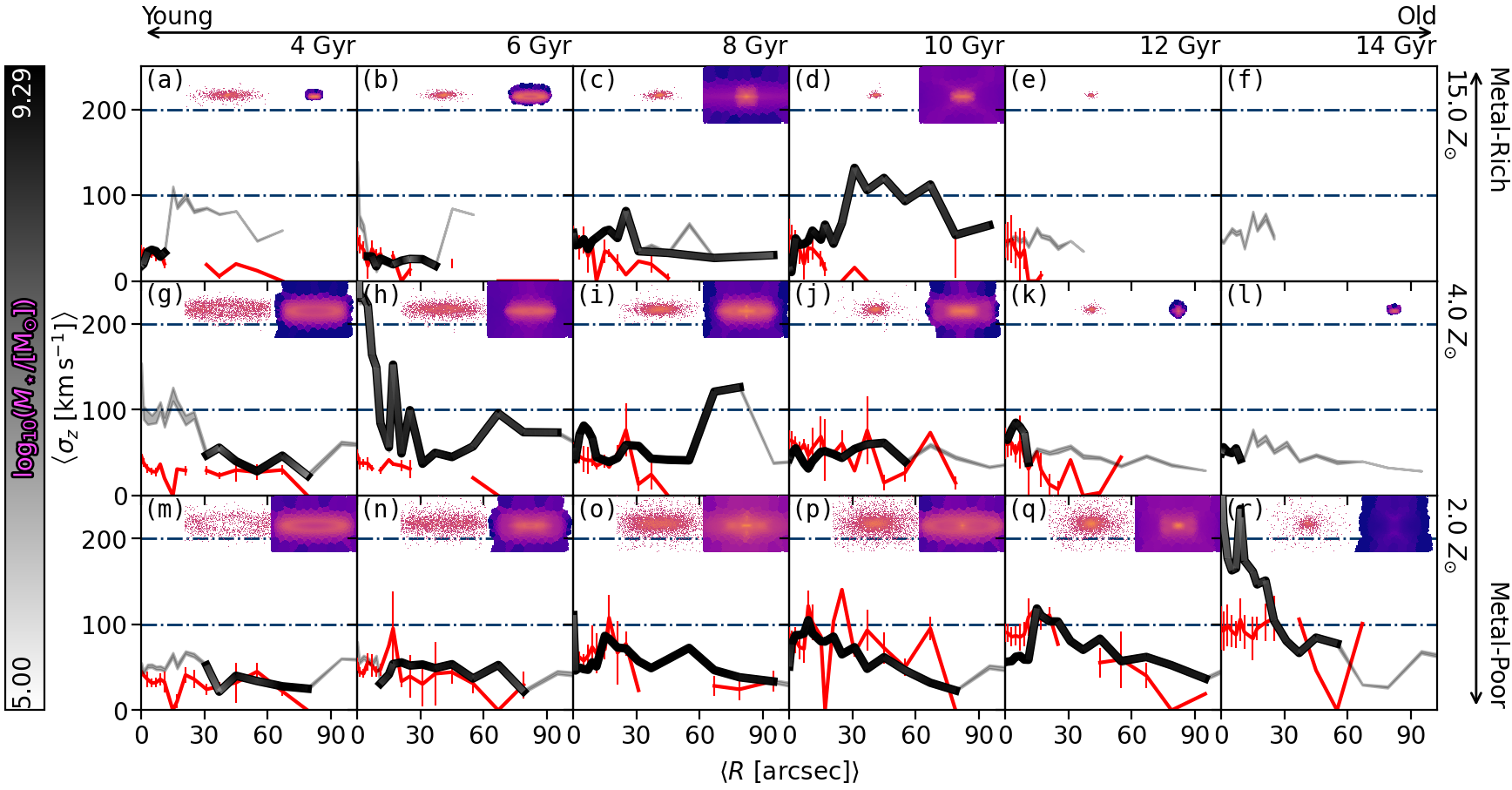}}
    \caption{Similar to \protect\cref{img:mfh153}, showing the assembly history for halo 5 ({\em top}) and halo 6 ({\em bottom}). In addition to the \(\sigma_z\) radial profile ({\em black/white line}) and surface brightness distribution ({\em upper right}) from the \shw\ model, the shaded regions correspond to the variations derived from \(100\) Monte Carlo fits to the stellar population maps. Each panel also contains the intrinsic \(\sigma_z\) radial profile ({\em red line}) and surface brightness distribution ({\em upper left}) from the simulations.}
    \label{img:mockMAH}
\end{figure*}
\begin{figure}
    \centerline{
        \includegraphics[width=\columnwidth]{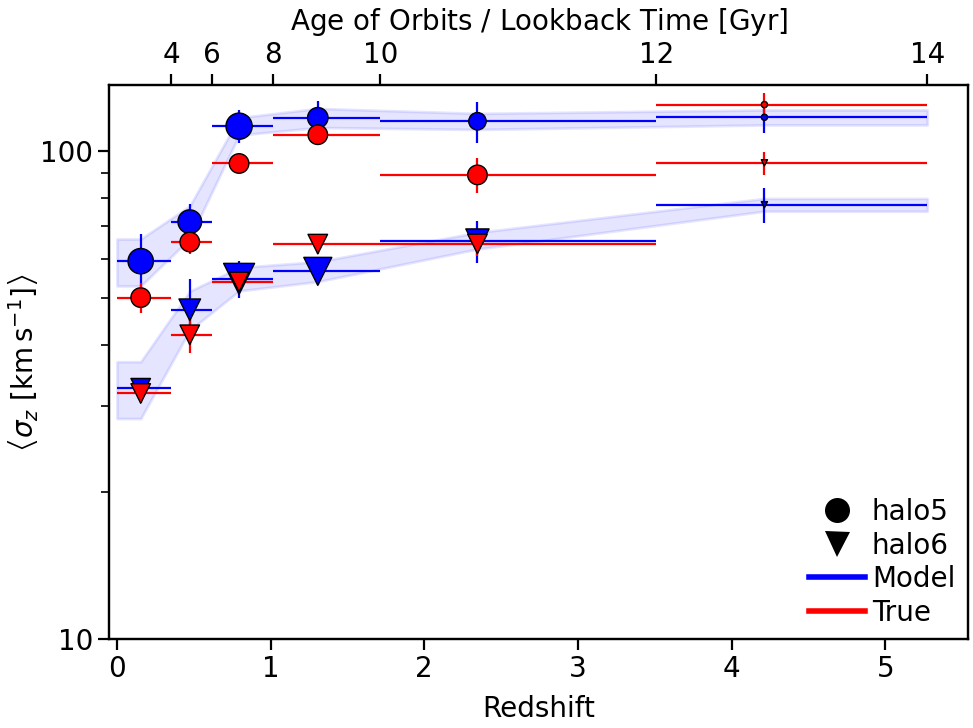}}
    \caption{Similar to \protect\cref{img:cosmoDisp}, showing the recovered ({\em blue}) and intrinsic ({\em red}) AVR for halos 5 and 6.}
    \label{img:mockCosmoDisp}
\end{figure}

The recovery of the assembly histories is presented in \cref{img:mockMAH}. It can be seen that the radial profiles of \(\sigma_z\) from the simulations often extend to larger radii than the corresponding profiles from the \shw\ model. This is due to the fact that each particle in the simulation contributes to its profile, while for the \shw\ model orbits are required to have spent some time in a particular region before being included. A single particle (or very few) in the simulations of a given \(t-\chemZH\) component may give the impression of having a larger extent compared to the \shw\ model, while actually contributing negligible mass. Therefore we emphasise that it is the common radial regions for each pair of profiles, and the overall spatial distribution of particles/orbits for each panel, that provides the best indication of the model's recovery as this is where the vast majority of the mass resides. In this regime, there is clearly point-to-point variation between the intrinsic and recovered radial profiles of velocity dispersion, which are noisy for some components. The agreement is worse for decreasing mass, which is expected. Nevertheless, it can be seen that the general trends are accurately recovered by the model, in particular the relative dynamics between the different stellar populations. The absolute quantitative agreement between the model and simulations is also good, though in some cases there appears to be additional discretisation effects in the model producing sharply-varying profiles in the low-mass panels, which is likely due to the binning in circularity space combined with the binning in the \(t-Z\) space. Reassuringly, the \ND{2} distributions are also well-matched by the model. Some mass is re-distributed to different bins compared to the simulations, which are typically the compact spheroidal structures. Such `boundary' effects are inevitable when imposing discrete binning on the data, as those components with values close to these boundaries may straddle one way or the other. The Monte Carlo simulations show similar behaviour, and in this way the shaded regions account for these boundary effects.
Finally, the recovery of the AVR are shown in \cref{img:mockCosmoDisp}. Once again, we see good qualitative and quantitative agreement.\par
We conclude that the model is sufficiently robust to be able to draw strong inferences about real galaxies. With regards to the main conclusions of this work, the relative behaviour of the chemo-dynamical populations is recovered well, where the model stellar AVR for both Auriga halos are consistent to within \(\sim 10\ \si{\kilo\metre\per\second}\) of their intrinsic values. We highlight that we believe this is an upper limit on account of our real data having higher spatial resolution, the corresponding models having higher orbit sampling, and all galaxies studied here having \(\theta > 80\si{\degree}\) \citep[where more edge-on projections are more reliable;][]{zhu2020}, which all work in favour of improving the reliability of this technique further. We are therefore confident that this method is able to recover not only projected average quantities like the \ND{2} stellar-population maps of \cref{img:mockMW}, but also quantitatively recover the underlying chemo-dynamic distributions, and that our main conclusions are robust against systematic effects.

\end{appendix}

\end{document}